\documentclass[12pt]{iopart}
\usepackage{iopams}
\usepackage{ mathrsfs, bbm, bm, graphicx, pdfsync, textcomp}
\usepackage[T1]{fontenc}
\usepackage{slashed}
\usepackage{comment}
\usepackage{axodraw4j}
\usepackage{pstricks}
\usepackage{color, caption}
\usepackage{subfig}

\usepackage{cite}
\newif \ifContLineOne
\newif \ifContLineTwo
\newif \ifContLineThree

\def \conC#1{\vbox{\ialign{##\crcr
\ifContLineThree \hrulefill\else \vphantom{\hrulefill}\fi \crcr
\noalign{\kern 3 . 2 pt \nointerlineskip }
\ifContLineTwo \hrulefill \else \vphantom{\hrulefill }\fi \crcr
\noalign{\kern 3 . 2 pt \nointerlineskip }
\ifContLineOne \hrulefill \else \vphantom{\hrulefill }\fi \crcr
\noalign{\nointerlineskip}
$\hfil \textstyle{\vbox to 14 pt{}#1}\hfil $\crcr }}}

\def \DrawLeg#1#2{\kern-.2 pt 
\dimen2 =#1 
\advance\dimen2 by 2pt 
\dimen3 = 1 0. 6 pt 
\dimen4 =3.6 pt 
\advance\dimen3 by -\dimen2
\multiply\dimen4 by #2
\advance\dimen3 by \dimen4
\raise \dimen2 \hbox{\vrule height \dimen3 width . 4 pt } 
\kern -.2 pt}

\newif\ifContLineOne
\newif\ifContLineTwo
\newif\ifContLineThree

\def\conC#1{\vbox{\ialign{##\crcr
  \ifContLineThree\hrulefill\else\vphantom{\hrulefill}\fi\crcr
  \noalign{\kern3.2pt\nointerlineskip}
  \ifContLineTwo\hrulefill\else\vphantom{\hrulefill}\fi\crcr
  \noalign{\kern3.2pt\nointerlineskip}
  \ifContLineOne\hrulefill\else\vphantom{\hrulefill}\fi\crcr
  \noalign{\nointerlineskip}
  $\hfil\textstyle{\vbox to 14pt{}#1}\hfil$\crcr}}}

\def\DrawLeg#1#2{
  \kern-.2pt              
  \dimen2 =#1             
  \advance\dimen2 by 2pt  
  \dimen3 = 10.6pt        
  \dimen4 =3.6pt          
  \advance\dimen3 by -\dimen2 
  \multiply\dimen4 by #2
  \advance\dimen3 by \dimen4
  \raise\dimen2 \hbox{\vrule height\dimen3 width .4pt} 
  \kern-.2pt}             

\def\begC#1#2{\setbox0 =\hbox{$\textstyle{#2}$}
  \dimen0=.5\wd0 \dimen1=\ht0
  \conC{\hskip\dimen0}
  \count255=#1
  \ifnum\count255 =1 \ContLineOnetrue\else
  \ifnum\count255 =2 \ContLineTwotrue\else
  \ifnum\count255 =3 \ContLineThreetrue\fi\fi\fi
  \DrawLeg{\dimen1}{\count255}
  \conC{\hskip\dimen0}
  \kern-\dimen0\kern-\dimen0 \box0}

\def\endC#1#2{\setbox0 =\hbox{$\textstyle{#2}$}
  \dimen0=.5\wd0 \dimen1=\ht0
  \conC{\hskip\dimen0}
  \count255=#1
  \ifnum\count255 =1 \ContLineOnefalse\else
  \ifnum\count255 =2 \ContLineTwofalse\else
  \ifnum\count255 =3 \ContLineThreefalse\fi\fi\fi
  \DrawLeg{\dimen1}{\count255}
  \conC{\hskip\dimen0}
  \kern-\dimen0\kern-\dimen0 \box0}

\begin{document}
\title[Non-Perturbative Dynamics in Massless \cal{QED}$_{2+1}$]{Non-Perturbative Dynamics, Pair Condensation, Confinement and Dynamical Masses in Massless \cal{QED}$_{2+1}$}
\author{Micha\"el Fanuel$^1$ and Jan Govaerts$^{1,2,}$}
\vspace{10pt}
\address{$^1$Centre for Cosmology, Particle Physics and Phenomenology (CP3),\\
Institut de Recherche en Math\'ematique et Physique,\\
Universit\'e catholique de Louvain, Chemin du Cyclotron 2, bte L7.01.01,\\
B-1348, Louvain-la-Neuve, Belgium}
\vspace{10pt}
\address{$^2$International Chair in Mathematical Physics\\
and Applications (ICMPA--UNESCO Chair),\\
University of Abomey--Calavi, 072 B. P. 50, Cotonou, Republic of Benin}
\ead{Michael.Fanuel@uclouvain.be, Jan.Govaerts@uclouvain.be}

\begin{abstract}
Quantum electrodynamics in three spacetime dimensions, with one massless fermion species, is studied using a non-perturbative variational approach. Quantization of the theory follows Dirac's Hamiltonian procedure, with a gauge invariant factorization of the physical degrees of freedom. Due to pair condensation in the vacuum state, the symmetry of parity is spontaneously broken. As a consequence, fermionic quasi-particles propagating in the condensate can be identified and are seen to possess a confining dynamical mass, while the propagating physical electromagnetic mode also acquires a non-vanishing dynamical mass. The issues of gauge invariance and confinement of the constituent fermions are carefully discussed.
\end{abstract}

\section{Brief overview and motivations}

The non-perturbative dynamics of gauge theories remains a challenging issue, in particular in the case of the strong coupling regime of quantum chromodynamics. Indeed, many techniques are still being developed, such as lattice gauge theories, and functional equations to unravel the question (for a review, see for example \cite{Alkofer:2010ue}).
A pioneer study by Polyakov \cite{Polyakov:1976fu} of compact \cal{QED}$_{2+1}$ in the absence of dynamical matter established the confinement of charges, and generated further analyses in lattice gauge theory \cite{Gopfert:1981er}. Furthermore, non-compact \cal{QED}$_{2+1}$ including dynamical fermions, whose massless version will be studied in this paper, attracted the interest of theoreticians for various reasons.

Undeniably, important features render this theory an interesting laboratory in order to develop techniques addressing non-perturbative dynamics. Namely, the excellent ultraviolet behaviour of perturbative \cal{QED}$_{2+1}$ is remarkable. Among the primary divergent diagrams of \cal{QED}$_{3+1}$, only the electron self-energy and the vacuum polarisation of \cal{QED}$_{2+1}$ are superficially one-loop divergent. Following from gauge invariance and a symmetric integration of the loop, both diagrams are actually finite in dimensional regularisation. In a renowned paper \cite{JackiwTempleton}, Jackiw and Templeton analysed the infrared divergences occuring in perturbation theory in \cal{QED}$_{2+1}$ with massless fermions, while the excellent behaviour of the theory in the UV is emphazised. Using a toy model treated non-perturbatively, these authors explain how the perturbative expansion in the coupling constant has to be completed by an expansion in logarithms of the coupling constant, while they expect also contributions which remain beyond the reach of perturbation theory. 

In analogy with \cal{QCD}$_{3+1}$, the question of spontaneous chiral symmetry breaking has also been raised in the context of \cal{QED}$_{2+1}$ with $N$ flavours. Chirality may be defined in $2+1$ dimensions by considering $4$-spinors, in a reducible representation of the Lorentz group. The analysis of the Schwinger-Dyson equations with various truncation schemes leads to a critical number of flavours, varying slightly according to the different authors (see for example \cite{Braun:2014wja}).

On the other hand, \cal{QED}$_{2+1}$ unexpectedly arised as an effective theory of recently discovered condensed matter models. Remarkably a two-flavour version of massless \cal{QED}$_{2+1}$ has been shown to describe well the low energy dynamics of graphene\footnote{In a two dimensional material, the electromagnetic interaction is not confined to the material, yielding a Coulomb potential $\propto 1/|\vec{p}|$ in the Fourier space.}. Due to the cristalline structure, the valence and conduction bands of graphene meet in two inequivalent conical points in the fundamental cell. At these ``Dirac points'', the dispersion relation can be linearized. As a result, the quasi-particles in the material are Dirac fermions \cite{Semenoff}. The Fermi velocity being small compared to the speed of light, the effective coupling constant in graphene is approximately $300$ times larger than in \cal{QED}. The upshot is that the traditional approach based on perturbation theory has to be questioned.

Strikingly, solid state physics can also effectively reproduce the dynamics of a ``undoubled'' Dirac fermion in $2+1$ dimensions.
More recently, the discovery of a new class of materials called ``topological insulators'' \cite{HasanKaneRevModPhys} has opened a new age in condensed matter physics.
Indeed, the surface of a 3D strong topological insulator \cite{FuKane} exihibits a peculiar behaviour, since it is possible that the Fermi energy  intersects a single ``Dirac point''. The result is that the effective quasi-particle dynamics can be described by a single Dirac field. \\
Recent studies were also conducted in order to describe high-temperature superconductivity with a model based on \cal{QED}$_{3}$, understood as an effective theory, as explained for example in the recent work of Ref. \cite{Bonnet:2011hh}.

The present work stands in direct continuity with a first approach to the quantisation without gauge fixing of a gauge theory, exposed in the paper \cite{Fanuel:2011ds} by the present authors, which provided new insights into the solution of the Schwinger model, namely massless  \cal{QED}$_{1+1}$. More precisely, the technique mainly relies on a factorisation of the physical degrees of freedom \cite{Bertrand:2007ja,Bertrand:2007ay} and considers the dynamics of fermions ``dressed'' by their electric field, as first introduced by Dirac \cite{Dirac:1955uv}. The dressing of physical charges was elaborated further in \cite{Lavelle:1995ty}, for instance, and was shown to greatly improve the soft dynamics \cite{Bagan:1997kg}.

In what follows, the analysis will mostly concern \cal{QED}$_{2+1}$, in the absence of a mass term for fermions, with only one flavour of electrons. In this setting, a bare fermion mass term would break parity. This parity preserving formulation exposed here does not include a Chern-Simons term neither which would provide a mass to the photon. Indeed, in $2+1$ dimensions, the Chern-Simons coefficient and the fermion mass are intricately related. At the perturbative level, a bare mass term for the fermion will radiatively induce an abelian Chern-Simons term and conversely \cite{TopologicallyMassive,Dunne:1998qy,SpiridonovTkachov}. The choice is made to restrict ourselves to a Lagrangian invariant under parity. By the way, this version of \cal{QED}$_{2+1}$ was considered in \cite{JackiwTempleton}, while in recent years the confining property, the dynamical mass, and related aspects have been investigated with success by Y. Hoshino within another framework relying on the study of the position space fermion propagator \cite{Hoshino:2004ns}. The work exposed in the present paper relies on a different and complementary approach.  Incidentally, a possible way to engineer a theory with massive fermions, while preserving parity, is to introduce an even number of fermion flavours, having in pairs opposite mass terms. However we will not pursue this possibility here.

Here is a brief summary of the results presented in this paper.
Section \ref{ClassicalHamiltonian2+1} deals with the classical formulation of the theory. Working with a factorized gauge symmetry, we are facing the particular case of the logarithmic confining electrostatic potential. The Fourier transform of the $x$-space potential is found to be a distribution. 
Within the Hamitonian framework, Section \ref{QuantisationQED2+1} deals with the quantisation of the theory, and the construction of a non-perturbative approximation. In order to look for a stable ground state, a fermionic coherent state, similar to the BCS superconducting vacuum state, is considered, inspired by previous works \cite{Finger:1979yt,Finger:1981gm,GovaertsThesis,Govaerts:1983ft} in \cal{QED}$_{3+1}$ and \cal{QCD}$_{3+1}$.
In Section \ref{SectionFermionCondensate}, we formulate an integral equation for the vacuum wave function from the requirement of energy minimization for this trial state. 
An approximate solution to the integral equation is found, inclusive of the effects of an infinite number of photon exchanges. The energy density of this condensate is lower than the energy density of the Fock state, so that the Fock state is expected to be unstable.
Spontaneous parity violation with only one fermion flavour is brought forth by the condensation, supporting a similar argument by Hoshino and Matsuyama \cite{Hoshino:1989ct,Matsuyama:2000bn}. 

By analysing in Section \ref{QuasiParticleHamiltonian} the dynamics of fermions in the condensate, quasi-particles interpreted as constituent fermions are identified and their dispersion relation is studied. The divergence of their energy at zero momentum is a signature for the confinement of dynamical charges, as confirmed in Section \ref{SectionEnergyPair}. Subsequently, a Green function interpretation of the results of the variational analysis is presented in Section \ref{SectionGreenFunctions}. Treating the residual interactions as perturbations, the analysis is in favour of a dynamical mass for the fermions.

Finally, the effect of the condensate on the electromagnetic sector is addressed in Section \ref{SectionCorrectionMagneticMode} showing that the physical propagating electromagnetic mode acquires a dynamical mass as well, while Section \ref{Conclusion2+1} is devoted to some conclusions.

\section{Classical Hamiltonian \cal{QED}$_{2+1}$ \label{ClassicalHamiltonian2+1}}

The analysis starts with the statement of the conventions chosen. In order to appropriately describe a single fermion flavour, Dirac matrices are chosen in terms of the Pauli matrices as follows: $\gamma^{0}=\sigma_{3}$ and $\gamma^{i}=\rmi\sigma_{i}$ for $i=1,2$, and satisfy the useful properties
\begin{eqnarray}
\Tr (\gamma_{\mu}\gamma_{\nu})=2\eta_{\mu\nu}, \quad \Tr (\gamma_{\mu}\gamma_{\nu}\gamma_{\rho})=-2\rmi\epsilon_{\mu\nu\rho},
\end{eqnarray}
where the totally anti-symmetric symbol is chosen so that $\epsilon_{012}=\epsilon^{012}=1$.
The mostly minus signature is chosen for the Minkowski metric, while an implicit choice of units is done such that $\hbar=c=1$. As for the dimensional specificities, in $D=3$ space-time dimensions, and in units of mass $M$ the gauge coupling constant $e$ has dimension $[e]=M^{1/2}$, while the gauge and matter fields have dimensions $[A_{\mu}]=M^{1/2}$ and $[\psi]=M^{1}$. 

\subsection{Classical Hamiltonian and the Green function}
The classical dynamics is given by the Lagrangian density 
\begin{eqnarray}
\mathcal{L}=-\frac{1}{4}F_{\mu\nu}F^{\mu\nu}+\frac{1}{2}\rmi\overline{\psi}\gamma^{\mu}(\partial_{\mu}+\rmi eA_{\mu})\psi-\frac{1}{2}\rmi(\overline{\partial_{\mu}+\rmi eA_{\mu})\psi}\gamma^{\mu}\psi.
\end{eqnarray}
We shall apply here a factorization of the local gauge transformations and gauge degrees of freedom, following closely the techniques explained in \cite{Fanuel:2011ds} for the case of the Schwinger model. In two space dimensions, the spatial gauge potential can be written\footnote{Henceforth, all latin indices are euclidian.} as the sum of a longitudinal and a transverse component 
\begin{eqnarray}
A_{i}(t,\vec{x})=\partial_{i}\phi(t,\vec{x})+\epsilon_{ij}\partial_{j}\Phi(t,\vec{x}),
\end{eqnarray}
where the scalar $\Phi$ is related the magnetic field through $\Delta \Phi=B$, so that $\Phi$ will be referred to as the ``magnetic mode''. Similarly, we also introduce the decomposition
\begin{eqnarray}
A_{0}(t,\vec{x})=a_{0}(t)+\partial_{i}\omega_{i}(t,\vec{x}).
\end{eqnarray}
The local gauge parameter may also be decomposed as the sum of its ``global'' (by which we mean throughout a space independent but yet possibly a time dependent gauge transformation parameter) and local components, $\alpha(t,\vec{x})=\beta_{0}(t)+\partial_{i}\beta_{i}(t,\vec{x})$. In order to factorize these local gauge transformations, the fermion field is ``dressed'', in a way completely analogous to that of reference \cite{Dirac:1955uv},
\begin{eqnarray}
\chi(t,\vec{x})=e^{\rmi e \phi(t,\vec{x})}\psi(t,\vec{x}),
\end{eqnarray}
so that the dressed fermion transforms, under gauge transformations of general parameter $\alpha(t,\vec{x})=\beta_{0}(t)+\partial_{i}\beta_{i}(t,\vec{x})$, only by a global (time dependent) phase change
\begin{eqnarray}
\chi(t,\vec{x})\to e^{-\rmi\beta_{0}(t)}\chi(t,\vec{x}).
\end{eqnarray}
Following the study of the Hamiltonian dynamics of constrained systems, as advocated by Dirac (see for example \cite{Govaerts:1991gd}), we give only a few details of the constrained analysis which is analogous to the one given in \cite{Fanuel:2011ds}. From the previous definitions, we obtain the Lagrangian action as a function of the new configuration space variables 
\begin{eqnarray*}
&S=\int dt&\Big\{-ea_{0}(t)\int_{S^1} \rmd x^{i}\chi^{\dagger} \chi+\int \rmd x^{i}\Big(\frac{1}{2}\rmi\chi^{\dagger}\partial_{0}\chi-\frac{1}{2}\rmi\partial_{0}\chi^{\dagger}\chi\\
&&+\frac{1}{2}\rmi\overline{\chi}\gamma^i\partial_{i}\chi-\frac{1}{2}\rmi\partial_{i}\overline{\chi}\gamma^i\chi-\frac{1}{2}(\partial_{0}\phi-\partial_{i}\omega_{i})\Delta(\partial_{0}\phi-\partial_{i}\omega_{i})\\
&&+e(\partial_{0}\phi-\partial_{i}\omega_{i})\chi^{\dagger}\chi-\frac{1}{2}\partial_{0}\Phi\Delta\partial_{0}\Phi-\frac{1}{2}\Phi\Delta^{2}\Phi\\
&&-e\epsilon^{ij}\partial_{j}\Phi\bar{\chi}\gamma^{i}\chi\Big)\Big\}.
\end{eqnarray*}
 In order to study the Hamiltonian structure, we identify the conjugate momenta
\begin{eqnarray*}
&\pi_{\Phi}&=\frac{\partial L_0}{\partial \dot{\Phi}}\\
&p^{0}&=\frac{\partial L_0}{\partial \dot{a}_0}=0,\\
&\pi^{i}&=\frac{\partial L_0}{\partial \dot{\omega}_{i}}=0\\
&\pi_{\phi}&=\frac{\partial L_0}{\partial \dot{\phi}}=-\bigtriangleup (\partial_{0}\phi-\partial_{i}\omega_{i})+e(\chi^{\dagger}\chi),\\
&\xi_{1}&=\frac{\partial L_0}{\partial \dot{\chi}}=-\frac{1}{2}\rmi\chi^\dagger,\\
&\xi_{2}&=\frac{\partial L_0}{\partial \dot{\chi}^\dagger}=-\frac{1}{2}\rmi\chi ,
\end{eqnarray*}
where we observe that the fermion field is already in Hamiltonian form. Subsequently, the constraint analysis can be performed in close analogy with \cite{Fanuel:2011ds}, while the first class constraints $p^{0}=0$ and $\pi^{i}=0$ can be solved. After this straightforward analysis, the equations of motion of the sector ($\phi$, $\pi_{\phi}$) can be used to reduce these phase space variables from the dynamics. Finally, we obtain the following Hamiltonian action
\begin{eqnarray}
S=\int \rmd t\Big\{\int\rmd^{2}x^{i} \Big[\partial_{0}\Phi\pi_{\Phi}+\frac{1}{2}\rmi\chi^{\dagger}\partial_{0}\chi-\frac{1}{2}\rmi\partial_{0}\chi^{\dagger}\chi\Big]-H\Big\}
\end{eqnarray}
where the classical expression of the Hamitonian is
\begin{eqnarray}
H=\int\rmd^{2} x^{i}\Big\{\mathcal{H}_{F}+\mathcal{H}_{\Phi}+\mathcal{H}_{\Phi\chi}\Big\},\label{HamiltonianClassical}
\end{eqnarray}
with the Hamiltonian densities
\begin{eqnarray}
&\mathcal{H}_{F}&=\frac{1}{2}\bar{\chi}(t,\vec{x})\gamma^{i}(-\rmi\partial_{i})\chi(t,\vec{x})+\frac{1}{2}\rmi\partial_{i}\bar{\chi}(t,\vec{x})\gamma^{i}\chi(t,\vec{x})\nonumber\\
& &-\frac{e^{2}}{2}(\chi^{\dagger}\chi)(t,\vec{x})\big[\Delta^{-1}(\chi^{\dagger}\chi)\big](t,\vec{x}),\\
&\mathcal{H}_{\Phi}&=-\frac{1}{2}\pi_{\Phi}(t,\vec{x})\big[\Delta^{-1}\pi_{\Phi}](t,\vec{x})+\frac{1}{2}(\Delta\Phi)^{2}(t,\vec{x}),\\
&\mathcal{H}_{\Phi\chi}&=e\epsilon^{ij}\partial_{j}\Phi(t,\vec{x})(\bar{\chi}\gamma^{i}\chi)(t,\vec{x}).\label{HPhiChi}
\end{eqnarray}
On account of the factorisation of local gauge transformations and gauge degrees of freedom, the dynamics is still constrained by the condition stemming from the time-dependent ``global'' gauge transformations with $\alpha(t)=\beta_{0}(t)$ which is analogous to the spatially integrated Gauss law,
\begin{eqnarray}
\int\rmd^{2}x^{i} \chi^{\dagger}(t,\vec{x})\chi(t,\vec{x})=0,
\end{eqnarray}
which is first class and generates the remaining global gauge transformations.
Examining more closely the terms in (\ref{HamiltonianClassical}), we observe that the Hamiltonian density $\mathcal{H}_{F}$ describes the dynamics of the fermion with its Coulomb interaction, while $\mathcal{H}_{\Phi}$ characterizes the dynamics of the magnetic mode sector. The Hamiltonian density $\mathcal{H}_{\Phi\chi}$ accounts for the interaction between the fermion current and the magnetic mode.

In order to understand the quantum theory, we first need to study the peculiarities of the Green function of the Laplacian in two spatial dimensions.
A peculiarity of this $2+1$-dimensional theory is that the Green function of the spatial Laplacian, conveniently expressed in $x$-space and verifying $\Delta G(\vec{x},\vec{y})=\delta^{(2)} (\vec{x}-\vec{y})$,
 is the tempered distribution defined by
\begin{eqnarray}
G(\vec{x},\vec{y})=\frac{1}{2\pi}\ln( \mu | \vec{x} - \vec{y} |),\label{Chap2Green2D}
\end{eqnarray}
where the mass scale $\mu>0$ is introduced for dimensional consistency. In classical electrostatics, this Green function is proportional to the electrostatic potential of a pointlike particle in two space dimensions. In three space dimensions, the electrostatic potential of an infinite charged wire would have a similar expression. The scale $\mu$ is therefore understood as parametrizing the possible choices for a ``zero of the potential'', and will be kept arbitrary in the sequel. When the potential tends to a constant at spatial infinity, it is allowed to choose this constant to be zero. On the contrary, because the logarithmic Coulomb potential is confining, the remaining gauge freedom $\mu$ has to be considered at all steps of the calculation. In $p$-space, the presence of $\mu$ can be interpreted as an infrared regulator, as we shall see.

Because the Green function is divergent at large as well as at small distances, we may expect to encounter also infrared divergences in the quantum formulation of the theory. We will pay special attention to the classical large distance divergence of the Green function. The inverse of the Laplacian is obtained by the convolution integral 
\begin{eqnarray}
(\Delta^{-1}f)(\vec{x})=\langle G(\vec{x},\cdot),f(\cdot)\rangle=\int\rmd y^{i}\frac{1}{2\pi}\ln(\mu |\vec{x}-\vec{y}|)f(\vec{y}).
\end{eqnarray}
Adding a constant to (\ref{Chap2Green2D}), amounts to redefining $\mu$ by a multiplicative constant.
For technical reasons, we should like to express the Green function in Fourier space. However the Fourier transform of the Green function is not a function, but rather a distribution. The naive expression for the Fourier transform, namely $\propto 1/|\vec{p}|^{2}$, would  indeed fail to converge in the infrared region. After a careful integration, one finds the identity
\begin{eqnarray}
\frac{1}{2\pi}\ln(\frac{e^{\gamma}}{2} \mu|\vec{x}-\vec{y}|)=G_{\epsilon}(\vec{x},\vec{y})-\frac{1}{2\pi}\ln (\epsilon/\mu),\label{Chap2GreenEpsilon}
\end{eqnarray}
where $\gamma$ is the Euler constant \footnote{This integration is performed with the help of $\int_{0}^{\infty} \ln y J_{1}(a y) \rmd y=(-1/a)( \ln(a/2)+\gamma)$, where $J_{1}$ is a Bessel function of the first kind.}. Here we have defined
\begin{eqnarray}
&G_{\epsilon}(\vec{x},\vec{y})=&\int_{|\vec{p}|<\epsilon} \frac{\rmd ^{2} p^{i}}{(2\pi)^{2}} \frac{-1}{|\vec{p}|^{2}}(e^{\rmi \vec{p}.(\vec{x}-\vec{y})}-1)\nonumber\\
&&+\int_{|\vec{p}|>\epsilon} \frac{\rmd ^{2} p^{i}}{(2\pi)^{2}}\frac{-1}{|\vec{p}|^{2}}e^{\rmi\vec{p}.(\vec{x}-\vec{y})},\label{Chap2Fourier2D}
\end{eqnarray}
where $\epsilon>0$ can take any value. The above is an exact result involving the arbitrary parameter $\epsilon$ playing the role of a cut-off  which makes the integral convergent close to the infrared singularity at $p=0$. The last definition (\ref{Chap2Fourier2D}) depends on the free parameter $\epsilon$ because we have $2\pi\partial_{\epsilon}G_{\epsilon}(\vec{x},\vec{y})=-1/\epsilon$. This dependence is, however, cancelled by the logarithmic term in (\ref{Chap2GreenEpsilon}).

\subsection{The Hadamard finite part}
In order to relate the discussion of the previous section to the mathematical theory of distributions, we will use here variables without physical dimensions. Restoring physical dimensions is straightforward.

In a renowned work \cite{Hadamard}, Hadamard introduced very useful generalized functions, among them the so-called Hadamard finite part $\mathcal{P}\frac{1}{x^{2}}$, which is related to the more popular Cauchy principal value $\mathcal{P}\frac{1}{x}$ by the ``weak'' derivative 
\begin{eqnarray}
\frac{\rmd}{\rmd x}\mathcal{P}\frac{1}{x}=-\mathcal{P}\frac{1}{x^{2}}.
\end{eqnarray}
This definition of the finite part is valid for functions of one variable, but it may be generalized to functions of two variables.
Following \cite{Vladimirov}, it is interesting to introduce here a two-dimensional version of the finite part of $1/x^{2}$, by defining its action on a test function $\phi$,
\begin{eqnarray}
(\mathcal{P} \frac{1}{|\vec{p}|^{2}},\phi)=\int_{|\vec{p}|<1} \rmd^{2} p^{i} \frac{\phi(\vec{p})-\phi(\vec{0})}{|\vec{p}|^{2}}+\int_{|p|>1} \rmd^{2} p^{i} \frac{\phi(\vec{p})}{|\vec{p}|^{2}},
\end{eqnarray}
where the presence of the value $1$ in the bounds of the integration domain is conventional. Let us denote the Fourier transform of the Green function of the Laplacian as $\mathcal{F}[G](\vec{p})$. We can now show that the generalized function $-\mathcal{P} \frac{1}{|\vec{p}|^{2}}$ is the ``generalized'' Fourier transform of the Green function, by proving that the Hadamard finite part solves $-|\vec{p}|^{2}\mathcal{F}[G](\vec{p})=1$. To do so we calculate

\begin{eqnarray}
&(|\vec{p}|^{2}\mathcal{P} \frac{1}{|\vec{p}|^{2}},\phi)&=(\mathcal{P} \frac{1}{|\vec{p}|^{2}},|\vec{p}|^{2}\phi)\nonumber\\
&&=\int_{|\vec{p}|<1} \rmd^{2} p^{i} \frac{|\vec{p}|^{2}\phi(\vec{p})-[|\vec{p}|^{2}\phi(\vec{p})]|_{0}}{|\vec{p}|^{2}}+\int_{|p|>1} \rmd^{2} p^{i} \frac{|\vec{p}|^{2}\phi(\vec{p})}{|\vec{p}|^{2}}\nonumber\\
&&=\int \rmd^{2} p^{i} \phi(\vec{p})=(1,\phi)
\end{eqnarray}
giving the solution $\mathcal{F}[G](\vec{p})=-\mathcal{P} \frac{1}{|\vec{p}|^{2}}$. This relation rephrases the results found in (\ref{Chap2GreenEpsilon}) and (\ref{Chap2Fourier2D}). Hence, the upshot is that the apparent IR divergent ``Coulomb'' propagator in $p$-space, proportional to $\frac{1}{|\vec{p}|^{2}}$ has not to be considered as a function. On the contrary, it should be understood as a generalized function, that is to say the Hadamard finite part $\mathcal{P}\frac{1}{|\vec{p}|^{2}}$. In the sequel we will see that in the absence of IR divergences, this last prescription reduces to the usual multiplication by the function $\frac{1}{|\vec{p}|^{2}}$.

Although instructive, the previous mathematical treatment could obscure one's physical intuition. It may be enlightening to relate the Hadamard finite part representation of the Fourier space Green function to a more usual treatment of the infrared singularities.
As is often done, a ``ad hoc'' mass term could be included for the photon to consider then the $p$-space Green function $\frac{1}{|\vec{p}|^{2}+\mu^{2}}$. The massless limit of the massive Green function could provide a more intuitive picture. \ref{HadamardAnnexe} explains how the Hadamard representation is recovered from the zero-mass limit of the massive Green function. 


\section{Quantum Hamiltonian and ordering prescription\label{QuantisationQED2+1}}

The careful and detailed definition of the Coulomb Green function will prove to be most relevant to the understanding of singularities in the quantum theory.
Given the classical formulation,  a quantum version can be formulated. Following the correspondence principle, classical (graded) Poisson brackets are replaced by quantum commutators or anti-commutators.
This formal quantization should be performed in both the fermionic and the bosonic sectors of the theory.  
\subsection{Magnetic sector}
As pointed out previously, the field $\Phi(t,\vec{x})$ is related to the magnetic field by the identity $\Delta\Phi=B$.
In order to quantise this sector, we decide to expand the magnetic mode and its momentum conjugate in terms of the plane wave Fock modes as follows, at the reference time $t=0$,
\begin{eqnarray}
&\Phi(0,\vec{x})&=\int\frac{\rmd^{2}k^{i}}{2\pi\sqrt{2}}\frac{-\rmi}{|\vec{k}|^{3/2}}\Big[\phi(\vec{k})e^{\rmi\vec{k}.\vec{x}}-\phi^{\dagger}(\vec{k})e^{-\rmi\vec{k}.\vec{x}}\Big],\\
&\pi_{\Phi}(0,\vec{x})&=\int\frac{\rmd^{2}k^{i}}{2\pi\sqrt{2}}(-|\vec{k}|^{3/2})\Big[\phi(\vec{k})e^{\rmi\vec{k}.\vec{x}}+\phi^{\dagger}(\vec{k})e^{-\rmi\vec{k}.\vec{x}}\Big], 
\end{eqnarray}
where the creators and annihilators satisfy $[\phi(\vec{\ell}),\phi^{\dagger}(\vec{k})]=\delta^{(2)}(\vec{\ell}-\vec{k})$, in order that fields obey the Heisenberg algebra $[\Phi(0,\vec{x}),\Pi_{\Phi}(0,\vec{y})]=\rmi\delta^{(2)}(\vec{x}-\vec{y})$. In a familiar way, the bosonic Fock algebra is represented in a Fock space, with the annihilators satisfying $\phi(\vec{\ell})|0\rangle=0$.  Since the quantisation procedure introduces ordering ambiguities, we decide to define the normal ordered form of a composite operator, in the magnetic sector, as the operator written with all $\phi^{\dagger}$'s to the left of all $\phi$'s.
Therefore, the normal ordered ``magnetic'' Hamiltonian, associated to a ``free'' field, 
\begin{eqnarray}
\hat{H}_{\Phi}=\int\rmd^{2} x^{i}:\Big\{-\frac{1}{2}\pi_{\Phi}(0,\vec{x})\big[\Delta^{-1}\pi_{\Phi}](0,\vec{x})+\frac{1}{2}(\Delta\Phi)^{2}(0,\vec{x})\Big\}:
\end{eqnarray}
may be expanded in modes as follows:
\begin{eqnarray}
\hat{H}_{\Phi}=\int\rmd^{2} k^{i}|\vec{k}|\phi^{\dagger}(\vec{k})\phi(\vec{k}).\label{FreeMagneticHamiltonian}
\end{eqnarray}
Treating $\hat{H}_{\Phi}$ as the free Hamiltonian and the other terms as interactions, considered in perturbation theory, we define the interaction picture field as
\begin{eqnarray}
\Phi_{I}(t,\vec{x})=e^{\rmi\hat{H}_{\Phi}t} \Phi(0,\vec{x}) e^{-\rmi\hat{H}_{\Phi}t}.
\end{eqnarray}
Using customary techniques, the free magnetic mode propagator, i.e. in absence of interaction, can be computed, producing the Feynman propagator
\begin{eqnarray}
\langle 0|T\Phi_{I}(x^{0},\vec{x})\Phi_{I}(0,\vec{0})|0\rangle
=\int \frac{\rmd^{3} k}{(2\pi)^{3}}\frac{e^{-\rmi k^{0}x^{0}+\rmi\vec{k}.\vec{x}}}{|\vec{k}|^{2}}\frac{\rmi}{(k^{0})^{2}-|\vec{k}|^{2}+\rmi\epsilon}.
\end{eqnarray}
The $p$-space propagator is illustrated by a curly line,
\begin{center}
\fcolorbox{white}{white}{
  \begin{picture}(65,16) (190,-152)
    \SetWidth{1.0}
    \SetColor{Black}
    \Vertex(191,-144){2}
    \Gluon(191,-144)(254,-144){1.5}{4}
    \Vertex(254,-144){2}
  \end{picture}
}
\end{center}
being a useful representation of the momentum space two-point function of the gauge invariant and physical magnetic mode.
Incidentally, after the elimination of the longitudinal gauge mode, the spatial gauge potential is $A_{T}^{i}=\epsilon^{ij}\partial_{j}\Phi$. Using this last identity and translational invariance, we recover the transverse photon propagator $D^{ij}(x^{0}-y^{0},\vec{x}-\vec{y})=\langle 0|TA_{T}^{i}(x^{0},\vec{x})A_{T}^{j}(y^{0},\vec{y})|0\rangle$ with

\begin{eqnarray}
D^{ij}(x^{0},\vec{x})=\rmi\int \frac{\rmd^{3} k}{(2\pi)^{3}}e^{-\rmi k^{0}x^{0}+\rmi\vec{k}.\vec{x}}\frac{\delta^{ij}-k^{i}k^{j}/\vec{k}^{2}}{(k^{0})^{2}-|\vec{k}|^{2}+\rmi\epsilon}
\end{eqnarray}
as follows from the identity $\epsilon^{im}k^{m}\epsilon^{jn}k^{n}=\vec{k}^{2} \delta^{ij}-k^{i}k^{j}$ (for a reference concerning Coulomb gauge \cal{QED}$_{2+1}$, see \cite{TopologicallyMassive}).
\subsection{Fermionic sector}
In order to quantise the fermion sector, the classical spinor field is expanded in the basis of solutions of the free Dirac equation.
The classical solutions to the Dirac equation in $2+1$ dimensions are constructed in terms of the spinors
\begin{eqnarray}
u(k^{\mu})=\left(
\begin{array}{c}
\frac{k^{2}+\rmi k^{1}}{\sqrt{k^{0}-m}}\\
\sqrt{k^{0}-m}
\end{array}\right),v(k^{\mu})=\left(
\begin{array}{c}
\frac{k^{2}+\rmi k^{1}}{\sqrt{k^{0}+m}}\\
\sqrt{k^{0}+m}
\end{array}\right),
\end{eqnarray}
normalized as $u^{\dagger}(k^{\mu})u(k^{\mu})=v^{\dagger}(k^{\mu})v(k^{\mu})=2k^{0}>0$ and where $k^{\mu}=(k^{0},\vec{k})$.
In the massless limit, the Dirac spinors $u(k^{\mu})=v(k^{\mu})$ are degenerate so that the mode expansions of the fields at $x^{\mu}=(x^{0},\vec{x})$ become
\begin{eqnarray}
&\chi(x^{\mu})=\int\frac{\rmd^{2} k^{i}}{2\pi\sqrt{ 2k^{0}}}\Big[b(\vec{k})e^{-\rmi k.x}+d^{\dagger}(\vec{k})e^{\rmi k.x}\Big]u(\vec{k}),\\
&\chi^{\dagger}(x^{\mu})=\int\frac{\rmd^{2} k^{i}}{2\pi\sqrt{ 2k^{0}}}\Big[b^{\dagger}(\vec{k})e^{\rmi k.x}+d(\vec{k})e^{-\rmi k.x}\Big]u^{\dagger}(\vec{k}),
\end{eqnarray}
 where the last two expressions have to be evaluated at $k^{0}=|\vec{k}|$, whereas $k.x= k^{0}x^{0}-\vec{k}.\vec{x}$ stands for the Minkowski inner product. Quantisation is performed at the reference time $x^{0}=0$. Following from the algebra of classical Dirac brackets, in the quantised theory the fermionic creators-annihilators have to verify $\{b(\vec{p}),b^{\dagger}(\vec{q})\}=\delta^{(2)}(\vec{p}-\vec{q})=\{d(\vec{p}),d^{\dagger}(\vec{q})\}$, while the fermionic Fock vacuum $|0\rangle$ is chosen to be annihilated by $b(\vec{p})$ and $d(\vec{p})$.
Let us consider an operator $AB$, bilinear in $b, d$ and their adjoints. Its contraction is defined to be,
\begin{eqnarray}
\begC1{A}\conC{ }\endC1{B}=\langle 0|AB|0\rangle
\end{eqnarray}
while its normal ordered form, where the creators are positioned to the left of all annihilators, is given by
 \begin{eqnarray}
:AB:=AB-\begC1{A}\conC{ }\endC1{B}.
\end{eqnarray}
With the help of these notations, the Hamiltonian operator is defined by a normal ordered form of the classical expression, where each charge density factor $\chi^{\dagger}\chi$ is also written in the normal order on its own:
\begin{eqnarray}
&\hat{H}=&\int\rmd^{2} x^{i}\ \frac{1}{2}:\bar{\chi}(t,\vec{x})\gamma^{i}(-\rmi\partial_{i})\chi(t,\vec{x}):+\frac{1}{2}:\rmi\partial_{i}\bar{\chi}(t,\vec{x})\gamma^{i}\chi(t,\vec{x}):\nonumber\\
& &+\hat{H}_{C}\label{Chap2QHamiltonian}
\end{eqnarray}
where
\begin{eqnarray}
\hat{H}_{C}=-\frac{e^{2}}{2}\int \rmd^{2} x^{i}\rmd^{2} y^{i}(:\chi^{\dagger}\chi:)(0,\vec{x})G(\vec{x},\vec{y})(:\chi^{\dagger}\chi:)(0,\vec{y}).\label{Chap2QHamiltonianCoulomb}
\end{eqnarray}
The Green function of the Laplacian $G(\vec{x},\vec{y})$ is given by (\ref{Chap2Green2D}). Gauss' law constraint, which involves the charge operator
 \begin{eqnarray}
\hat{Q}=\int\rmd^{2} x^{i}:\chi^{\dagger}(0,\vec{x}) \chi(0,\vec{x}) :,
\end{eqnarray}
annihilates the physical, i.e. gauge invariant quantum states, $\hat{Q}|\rm{phys}\rangle=0$, that is to say, the physical states should contain an equal number of fermions and anti-fermions, so that these states are electrically neutral.
This constraint may be connected with the problem of the divergences at large distances which is a typical concern in $2+1$ dimensional gauge theories. Let us explain how with an elementary argument. It is noteworthy that the classical electrostatic energy of a single pointlike charge is infrared divergent due to the logarithmic behaviour of the Green function. However, the electrostatic potential of a system made of two opposite pointlike charges is well behaved at large distances, because it is proportional to
 \begin{eqnarray}
\ln \mu |\vec{x}-\vec{x}_{1}|-\ln \mu |\vec{x}-\vec{x}_{2}|= \ln\frac{|\vec{x}-\vec{x}_{1}|}{|\vec{x}-\vec{x}_{2}|},
\end{eqnarray}
where $\vec{x}_{1}$ and $\vec{x}_{2}$ are the positions of the two opposite charges.
This classical argument strongly suggests that gauge invariant states should not suffer difficulties in the infrared region.
Accordingly, when $\hat{H}_{F}$ acts on a gauge invariant state, namely a state with a vanishing total charge, the result is not affected by the transformation $G(\vec{x},\vec{y})\to G(\vec{x},\vec{y})+\rm{cst}$, given the specific ordering of the charge density operators in the Coulomb Hamiltonian.

Thus, when we consider states containing an equal number of particles and anti-particles, we may simply substitute the naive expression for the Green function
\begin{eqnarray}
G(\vec{x},\vec{y})=\int_{(\infty)}\frac{\rmd ^{2} p^{i}}{(2\pi)^{2}}\frac{-1}{|\vec{p}|^{2}}e^{\rmi\vec{p}.(\vec{x}-\vec{y})},\label{Chap2NaiveGreenFunction}
\end{eqnarray}
apparently infrared divergent, in the formula for the quantum Hamiltonian $\hat{H}_{F}$. We may expect that no gauge dependence will occur due to the specific ordering prescription, provided that $\hat{H}_{F}$ acts on physical states. However this will not be true in the case of a single charged particle or anti-particle, as will be seen in the next Section.

\section{Fermion condensate in massless \cal{QED}$_{2+1}$\label{SectionFermionCondensate}}

Because a non trivial vacuum structure is expected from the classical features of the theory, we would like to investigate the possibility of a pair condensation mechanism in the vacuum. The approach followed here puts forward an expression of a trial state which is likely to provide a satisfactory approximation of the exact vacuum state. The developements are somehow inspired by the microscopic theory of low temperature superconductivity. We will try to argue that the choice is sufficiently flexible to provide a consistent approximation of the non-perturbative nature of the vacuum state. The freedom introduced by the trial state is associated to a ``wave function'' which is to be determined through a procedure of minimization of the total energy, in the presence of the Coulomb interaction. Interestingly, a very similar variational procedure, non explicitely Lorentz covariant, was very recently undertaken by Reinhardt {\it et al} in the case of Hamiltonian \cal{QCD}$_{3+1}$ in the Coulomb gauge \cite{PakReinhardt,Pak:2013uba}, opening the door to a novel approach.
This ``Hartree-Fock'' procedure has the avantage to provide a consistent framework to the approximation.

By the way, a different strategy to probe the non-perturbative effects  could rely on the functional formulation of quantum field theory. From this point of view, the problem would be to find a solution to the Schwinger-Dyson equations, with a specific truncation scheme and gauge fixing. Although these ideas might seem unrelated, we show that the problem to find a wave function minimizing the energy gives rise to an integral equation which can be formulated as Schwinger-Dyson equation for the fermion propagator. 

Inspired by the techniques developed in \cite{Finger:1979yt,Finger:1981gm,GovaertsThesis,Govaerts:1983ft}, which resulted in a successful description of non-perturbative properties of the pion \cite{Govaerts:1983ir, Finger:1980dw} and in a close analogy with the Bardeen-Cooper-Schrieffer ground state of a superconductor, we now introduce the coherent superposition
\begin{eqnarray}
|\Psi\rangle=\frac{1}{N(\Psi)}\exp[ -\int \rmd^{2} x^{i} \rmd^{2} y^{i}\ \tilde{\Psi}(|\vec{x}-\vec{y}|):\bar{\chi}(\vec{x})\chi(\vec{y}):]|0\rangle,
\end{eqnarray}
where $\tilde{\Psi}(|\vec{x}|)$ is a function describing the distribution in space of condensate pairs. Because of its convenience, it is advantageous to write the previous definition in momentum space. To do so, we perform a Fourier transform and find the expression
\begin{eqnarray}
|\Psi\rangle=\frac{1}{N(\Psi)}\exp \int \rmd^{2} p^{i}\ \Psi(|\vec{p}|)b^{\dagger}(\vec{p})d^{\dagger}(-\vec{p})|0\rangle,\label{Chap2CoherentState}
\end{eqnarray}
containing an arbitrary number of fermion/anti-fermion pairs of opposite momenta. Accordingly, it is guaranteed that the wave function is invariant under the spatial translations. The associated dimensionless wave function in momentum space $\Psi(p)=\Psi(|\vec{p}|)$ is chosen to be invariant under rotations in the plane. Because this function is complex valued, we can express it as the product of a modulus and a phase: $\Psi(p)=|\Psi(p)|\exp\rmi\phi(p)$ where $p=|\vec{p}|$.
The purpose of our analysis is to determine if the dynamics triggers a pair condensate, whose profile is described by the wave function $\Psi(p)$ in $p$-space. As a means to compute the normalization of the trial state, the integral over the momenta may be discretized, allowing to express the exponential as an infinite product. The normalisation of each of these factors may then be calculated individually and the continuum limit be taken subsequently. For the sake of completeness, the normalization of the coherent superposition of pairs
\begin{eqnarray}
N(\Psi)=\prod_{p^{i}}\sqrt{1+|\Psi(p)|^{2}},
\end{eqnarray}
may be computed, the continuous product being approximated by a discretization of the momentum space into a lattice.
For further use, let us define the functions of $p=|\vec{p}|$
 \begin{eqnarray}
\alpha(p)=\frac{1}{\sqrt{1+|\Psi(p)|^{2}}}, \quad \beta(p)=\frac{\Psi(p)}{\sqrt{1+|\Psi(p)|^{2}}},
\end{eqnarray}
which can be associated to an angle $\Theta(p)$ defined by the relations $\cos \Theta(p)= \alpha(p)$ and $\sin \Theta(p)= |\beta(p)|$.
Consequently, the trial state (\ref{Chap2CoherentState}) may be formulated as a product of normalized factors
\begin{eqnarray}
|\Psi\rangle=\prod_{p^{i}} \Big[\alpha(p)+ \beta(p)b^{\dagger}(\vec{p})d^{\dagger}(-\vec{p})\Big]|0\rangle.
\end{eqnarray}
These definitions allow to better interpret the trial state as a fermionic ``coherent state''. In order to investigate its content in terms of fermionic components, we naturally remark now that the following identities:
\begin{eqnarray}
b(\vec{p})|\Psi\rangle=\Psi(p)d^{\dagger}(-\vec{p})|\Psi\rangle, \quad d(-\vec{p})|\Psi\rangle=-\Psi(p)b^{\dagger}(\vec{p})|\Psi\rangle,
\end{eqnarray}
are somehow reminiscient of the property of the canonical coherent states, which are eigenstates of the annihilation operator.
This property enjoins us to define a Bogoliubov transformation of the creators and annihilators
\begin{eqnarray}
B(\vec{p})&=\alpha(p)b(\vec{p})-\beta(p)d^{\dagger}(-\vec{p}),\\
B^{\dagger}(\vec{p})&=\alpha(p)b^{\dagger}(\vec{p})-\beta^{*}(p)d(-\vec{p}),\\
D(-\vec{p})&=\alpha(p)d(-\vec{p})+\beta(p)b^{\dagger}(\vec{p}),\\
D^{\dagger}(-\vec{p})&=\alpha(p)d^{\dagger}(-\vec{p})+\beta^{*}(p)b(\vec{p}),
\end{eqnarray}
which verify $B(\vec{p})|\Psi\rangle=0=D(-\vec{p})|\Psi\rangle$ and satisfy the Fock algebra $\{B(\vec{p}),B^{\dagger}(\vec{q})\}=\delta^{(2)}(\vec{p}-\vec{q})=\{D(-\vec{p}),D^{\dagger}(-\vec{q})\}$ while all other anti-commutators vanish.
In a similar fashion, the inverse relations are provided by
\begin{eqnarray}
b(\vec{p})&=\alpha(p)B(\vec{p})+\beta(p)D^{\dagger}(-\vec{p}),\\
b^{\dagger}(\vec{p})&=\alpha(p)B^{\dagger}(\vec{p})+\beta^{*}(p)D(-\vec{p}),\\
d(-\vec{p})&=\alpha(p)D(-\vec{p})-\beta(p)B^{\dagger}(\vec{p}),\\
d^{\dagger}(-\vec{p})&=\alpha(p)D^{\dagger}(-\vec{p})-\beta^{*}(p)B(\vec{p}).
\end{eqnarray}
Because the states created by $B^{\dagger}$ and $D^{\dagger}$ carry the same electric charge as the ones created by $b^{\dagger}$ and $d^{\dagger}$ and diagonalize the fermionic Hamiltonian $H_{F}$ up to some residual Coulomb interactions, the former states can be regarded as physical fermionic particles excited over the condensate. Consequently it is useful to define a new ordering prescription associated to the condensate $|\Psi\rangle$ of any operator $\hat{O}$, to be denoted by $: \hat{O} :_{\Psi}$, such that all $B^{\dagger}$ and $D^{\dagger}$ operators are positioned to the left of all $B$ and $D$ operators.
Technical tools developed in \cite{Finger:1981gm} can simplify the computations dramatically, as we shall outline briefly.

Considering a bilinear operator $AB$ in these fermionic creation and annihilation operators, one may change the ordering prescription thanks to the formula
\begin{eqnarray}
:AB: \ =\ :AB:_{\Psi}+\widehat{A B}, \quad \widehat{AB}= \langle \Psi|AB|\Psi\rangle-\langle 0|AB|0\rangle\label{Chap2WideHatDefinition}
\end{eqnarray}
which will be used in the sequel in order to calculate the necessary matrix elements. Given the definition of the Bogoliubov operators, the mode expansions of the fermionic fields at $x^{\mu}=(0,\vec{x})$ are modified.  Thus, a substitution gives readily the following expansions:
\begin{eqnarray}
&\chi(0,\vec{x})=\int\frac{\rmd^{2} k^{i}}{2\pi\sqrt{ 2k^{0}}}[B(\vec{k})N_{1}(k)u(\vec{k})+D^{\dagger}(-\vec{k})N_{2}(k)u(-\vec{k})]e^{\rmi \vec{k}.\vec{x}},\nonumber\\
&\chi^{\dagger}(0,\vec{x})=\int\frac{\rmd^{2} k^{i}}{2\pi\sqrt{ 2k^{0}}}[B^{\dagger}(\vec{k})u^{\dagger}(\vec{k})N^{\dagger}_{1}(k)+D(-\vec{k})u^{\dagger}(-\vec{k})N^{\dagger}_{2}(k)]e^{-\rmi \vec{k}.\vec{x}}.\nonumber
\end{eqnarray}
For simplicity, the following matrices, whose definition are specific to the representation chosen for the Dirac matrices,
\begin{eqnarray}
&N_{1}(k)=\alpha(k)+\beta^{*}(k)\gamma^{0}, \quad &N_{1}^{\dagger}(k)=\alpha(k)+\beta(k)\gamma^{0},\\
&N_{2}(k)=\alpha(k)-\beta(k)\gamma^{0},\quad &N_{2}^{\dagger}(k)=\alpha(k)-\beta^{*}(k)\gamma^{0},
\end{eqnarray}
are introduced.
Being equipped with suitable tools, we may now envisage to compute the average kinetic and interaction energy of the state $|\Psi\rangle$. Since we work in a space of infinite volume the most favourable state will be the one minimizing the energy per unit volume. More precisely, we would like to calculate the energy density of the coherent state (\ref{Chap2CoherentState}), as given by
\begin{eqnarray}
 E=\frac{\langle\Psi|\hat{H}_{F}|\Psi\rangle}{(2\pi)^{2}\delta^{(2)}_{(p)}(0)},
\end{eqnarray}
where $(2\pi)^{2}\delta^{(2)}_{(p)}(0)$ is the  spatial ``volume'' and $\delta^{(2)}_{(p)}(0)$ the Dirac delta function in momentum space, in order to find the best wave function $\Psi(p)$ minimizing this ratio.
The computation of the energy density of the condensate requires the use of the Wick theorem to evaluate the product of normal ordered factors appearing in the Coulomb Hamiltonian
\begin{eqnarray}
&:\chi^{\dagger}_{\alpha}(\vec{x})\chi_{\alpha}(\vec{x}):G(\vec{x},\vec{y}):\chi^{\dagger}_{\beta}(\vec{y})\chi_{\beta}(\vec{y}):=& \label{Chap2OrderingHC}\\
&:\chi^{\dagger}_{\alpha}(\vec{x})\chi_{\alpha}(\vec{x})G(\vec{x},\vec{y})\chi^{\dagger}_{\beta}(\vec{y})\chi_{\beta}(\vec{y}):+:\chi^{\dagger}_{\alpha}(\vec{x})\chi_{\alpha}\begC1{(\vec{x})}\conC{G(\vec{x},\vec{y})\chi^{\dagger}_{\beta}}\endC1{(\vec{y})}\chi_{\beta}(\vec{y}):&\nonumber\\
&+:\chi_{\alpha}(\vec{x})\chi^{\dagger}_{\alpha}\begC1{(\vec{x})}\conC{G(\vec{x},\vec{y})\chi_{\beta}}\endC1{(\vec{y})}\chi^{\dagger}_{\beta}(\vec{y}):+\chi^{\dagger}_{\alpha}(\begC2{\vec{x}}\conC{ ) \chi_{\alpha}(}\begC1{\vec{x}}\conC{  ) G(\vec{x},\vec{y})\chi^{\dagger}_{\beta}(}\endC1{\vec{y}}\conC{)\chi_{\beta}( }\endC2{\vec{y}})&\nonumber
\end{eqnarray}
where the fields have been implicitly expressed at $x^{0}=0=y^{0}$.
It is necessary to compute the mean value of the last operator in the vacuum state $|\Psi\rangle$. To do so, following \cite{Finger:1979yt,Finger:1980dw}, we may take advantage of the newly defined ordering prescription and express these same operators in the order $: \ :_{\Psi}$, so that the calculation of the matrix elements is made simpler. Making use of the relation (\ref{Chap2WideHatDefinition}), we find
\begin{eqnarray}
&\langle\Psi|:\chi^{\dagger}_{\alpha}(0,\vec{x})\chi_{\beta}(0,\vec{y}):|\Psi\rangle=\widehat{\chi^{\dagger}_{\alpha}(0,\vec{x})\chi_{\beta}(0,\vec{y})}\\
&\langle\Psi|:\chi_{\alpha}(0,\vec{x})\chi^{\dagger}_{\beta}(0,\vec{y}):|\Psi\rangle=\widehat{\chi_{\alpha}(0,\vec{x})\chi^{\dagger}_{\beta}(0,\vec{y})}
\end{eqnarray}
and
\begin{eqnarray}
&\langle\Psi|:\chi^{\dagger}_{\alpha}(0,\vec{x})\chi_{\alpha}(0,\vec{x})\chi^{\dagger}_{\beta}(0,\vec{y})\chi_{\beta}(0,\vec{y}):|\Psi\rangle\nonumber\\
&=\widehat{\chi^{\dagger}_{\alpha}(0,\vec{x})\chi_{\beta}(0,\vec{y})}\quad\widehat{\chi_{\alpha}(0,\vec{x})\chi^{\dagger}_{\beta}(0,\vec{y})},
\end{eqnarray}
where $\alpha$ and $\beta$ denote the spinor components. For conciseness, useful formulas to calculate the above expressions can be found in  \ref{MatrixElements}. Theses results lead to the average energy 
\begin{eqnarray}
\langle\Psi|\hat{H}_{F}|\Psi\rangle=(2\pi)^{2}\delta^{(2)}_{(p)}(0)\int \frac{\rmd^{2}k^{i}}{(2\pi)^{2}}2|\vec{k}|\frac{|\Psi(k)|^{2}}{1+|\Psi(k)|^{2}}+ \langle\Psi|\hat{H}_{C}|\Psi\rangle,\label{Chap2EnergyDensity}
\end{eqnarray}
where the infrared finite mean interaction energy of the condensate is\footnote{A term proportional to $\hat{\ell}\times\hat{k}$ was omitted in this expression. The reason is that it was shown to vanish after the integral over the relative angle between $\vec{k}$ and $\vec{l}$.}
\begin{eqnarray}
& \langle\Psi|\hat{H}_{C}|\Psi\rangle=\nonumber\\
&-\frac{e^{2}}{2}\int \rmd^{2}x^{i}\rmd^{2}y^{i}\ \langle\Psi|:\chi^{\dagger}_{\alpha}(0,\vec{x})\chi_{\alpha}(0,\vec{x}):G(\vec{x},\vec{y}):\chi^{\dagger}_{\beta}(0,\vec{y})\chi_{\beta}(0,\vec{y}):|\Psi\rangle\nonumber\\
&=-\frac{e^{2}}{2}(2\pi)^{2}\delta^{(2)}_{(p)}(0)\int \frac{\rmd^{2}k^{i}\rmd^{2}\ell^{i}}{(2\pi)^{4}}\frac{-1}{(\vec{\ell}-\vec{k})^{2}}\Big\{\frac{1}{(1+|\Psi(k)|^{2})(1+|\Psi(\ell)|^{2})}\times\nonumber\\
& \times\Big[-2|\Psi(k)||\Psi(\ell)| \cos \phi(\ell)\cos \phi(k)\nonumber\\
&+\hat{\ell}.\hat{k}\Big(|\Psi(\ell)|^{2}+|\Psi(k)|^{2}-2|\Psi(\ell)||\Psi(k)|\sin \phi(\ell)\sin \phi(k)\Big)\Big]\nonumber\\
&+\frac{1}{2}-\frac{1}{2}\hat{\ell}.\hat{k}\Big\}.\label{Chap2MeanCoulombEnergy}
\end{eqnarray}
The expression of this mean interaction energy deserves some comments, because its finiteness is not self-evident.
Indeed, the last line of (\ref{Chap2MeanCoulombEnergy}), involving the factor of $\frac{1}{2}-\frac{1}{2}\hat{\ell}.\hat{k}$ is an infinite constant, corresponding to the term completely contracted in the last line of  (\ref{Chap2OrderingHC}) and which may be understood as a quantum fluctuation of the vacuum energy. It is divergent in the ultraviolet but not in the infrared as one can see from the limit $\vec{k}\to\vec{\ell}$, so that we choose to regulate it by introducing a momentum cut-off $\Lambda>0$,
\begin{eqnarray}
-\frac{e^{2}}{2}(2\pi)^{2}\delta^{(2)}_{(p)}(0)\int_{|\vec{k}|<\Lambda}\frac{\rmd^{2}k^{i}}{(2\pi)^{2}}\int_{|\vec{\ell}|<\Lambda}\frac{\rmd^{2}\ell^{i}}{(2\pi)^{2}} \frac{-1}{(\vec{\ell}-\vec{k})^{2}}(\frac{1}{2}-\frac{1}{2}\hat{\ell}.\hat{k}),\label{Chap2InfiniteConstant}
\end{eqnarray}
In presence of the regulator and since this contribution is independent of the condensate wave function $\Psi(p)$, we can safely subtract (\ref{Chap2InfiniteConstant}) from the Hamiltonian. This contribution is proportional to the bubble diagram
\begin{center}
\fcolorbox{white}{white}{
  \begin{picture}(60,59) (265,-83)
    \SetWidth{1.0}
    \SetColor{Black}
    \Arc[arrow,arrowpos=0.5,arrowlength=5,arrowwidth=2,arrowinset=0.2](294,-54)(28,180,540)
    \Photon(294,-26)(294,-82){5.5}{6}
    \Vertex(294,-26){2}
    \Vertex(294,-82){2}
  \end{picture}
}
\end{center}
where the exact meaning of this pictorial representation is given in terms of the Feynman rules listed in \ref{FeynmanRulesAnnexe}. 

Regarding the other terms in (\ref{Chap2MeanCoulombEnergy}), the apparent singularity of the integral at $\vec{k}=\vec{\ell}$, where a denominator vanishes, is resolved because the denominator appropriately goes to zero at the same time. The infrared finiteness of the mean Coulomb energy and its independence of the parameter $\mu$ are specifically due to the choice of ordering prescription in the definition of the Coulomb interaction, which is crucial. 

As it happens, the mean energy depends on both the modulus and the phase of the condensate wave function.
However, simple considerations about the interaction energy can provide information about the influence of the phase of the wave function on the magnitude of the interaction.
In order to minimize the energy density, we would like to make the interaction energy (\ref{Chap2MeanCoulombEnergy}) as negative as possible. A possibility is to require, separately, a stationary variation with respect to the phase and to the modulus of the wave function. We may first consider to choose the optimal phase of the condensate $\phi(p)$ to minimize the Coulomb energy. Varying $\langle\Psi|\hat{H}_{F}|\Psi\rangle$ with respect to $\phi(p)$, requires to take simply $\sin \phi(p)=0$ or $\cos \phi(p)=0$ for any $p>0$. Examining (\ref{Chap2MeanCoulombEnergy}), we notice that, because $\hat{\ell}.\hat{k}\leq 1$, the best choice is to maximize $\cos \phi(p)$, so that we take $ \phi(p)=0$, leading to a real wave function for the fermion condensate. Consequently, we decide to write in the sequel  $\Psi(p)=|\Psi(p)|$ to simplify the expressions.
\subsection{Integral equation}
Having formulated the expression of the expected energy density of the condensate, a necessary condition for finding an extremum of that quantity is given by the stationary variation of the energy density 
\begin{eqnarray}
 \frac{\delta}{\delta\Psi(p)}\frac{\langle\Psi|\hat{H}_{F}|\Psi\rangle}{(2\pi)^{2}\delta^{(2)}_{(p)}(0)}=0,
\end{eqnarray}
with respect to the wave function $\Psi(p)$. Dealing with the functional derivative in the case $p\neq 0$, the resulting nonlinear integral equation reads
\begin{eqnarray}
&p\Psi(p)=\frac{e^{2}}{8\pi^{2}}\int\frac{ \rmd^{2}q^{i} }{(\vec{q}-\vec{p})^{2}}[(1-\Psi(p)^{2})\frac{\Psi(q)}{1+\Psi(q)^{2}}+\hat{q}.\hat{p}\ \Psi(p)\frac{\Psi(q)^{2}-1}{\Psi(q)^{2}+1}].\label{Chap2IntegralEquation}
\end{eqnarray}
Owing to the invariance of the wave function under spatial rotations, the angular integral may be performed explicitly, with the help of formulas given in \ref{UsefulIntegrals}, so that the integral equation simplifies to
\begin{eqnarray}
&p\Psi(p)=&\alpha\int_{0}^{+\infty}\rmd q\Big[q\frac{1-\Psi(p)^{2}}{|p^{2}-q^{2}|}\frac{\Psi(q)}{1+\Psi(q)^{2}}\nonumber\\
&&\qquad\quad\quad+\frac{\Psi(p)}{2p} (-1+\frac{p^{2}+q^{2}}{|p^{2}-q^{2}|})\frac{\Psi(q)^{2}-1}{\Psi(q)^{2}+1}\Big].\label{Chap2DimensionalIntegralEquation}
\end{eqnarray}
where $\alpha=e^{2}/4\pi$. The non-perturbative features of the modelled phenomenon are reflected by the nonlinearity of the integral equation.
Notably, the integration converges in a neighbourhood of $q=p$ thanks to a cancellation of the two terms in the {\it rhs} of (\ref{Chap2DimensionalIntegralEquation}). The reason for the convergence at $q=p$ finds its origin in the choice of ordering prescription made for the Coulomb Hamiltonian.
Although obtaining an explicit analytical solution of the equation may be arduous, a property of the solution can be found without effort.
Actually, one may readily guess that, in order to ensure the convergence of the integral in the limit $p\to 0$, the wave function should verify $\Psi(0)=1$. The solution of the linearized equation is expected to have a very different behaviour close to $p=0$. The mathematical literature dealing with integral equations does not provide a suitable analytic method to find a solution to this kind of very non-linear equation with a singular kernel. As a consequence, we shall look for a numerical solution.

A possible concern about the integral equation could be the existence of solutions as the value of the coupling constant varies. To discuss the dependence on the parameter $\alpha$, one can try to understand how the equation depends on the typical scale of the problem.
In fact, it is possible to express the integral equation in terms of dimensionless variables, using $x=p/\alpha$ and $y=q/\alpha$,
\begin{eqnarray}
&x\psi(x)=&\int_{0}^{+\infty}\rmd y\Big[y\frac{1-\psi(x)^{2}}{|x^{2}-y^{2}|}\frac{\psi(y)}{1+\psi(y)^{2}}\nonumber\\
&&\qquad\quad\quad+\frac{\psi(x)}{2x} (-1+\frac{x^{2}+y^{2}}{|x^{2}-y^{2}|})\frac{\psi(y)^{2}-1}{\psi(y)^{2}+1}\Big],\label{Chap2AdimensionalIntegralEquation}
\end{eqnarray}
where, in terms of the wave function appearing in (\ref{Chap2IntegralEquation}), $\psi(x)=\Psi(\alpha x)$. The conclusion is that, whatever the value of $\alpha$, we have only one  equation to solve, which does not depend on $\alpha$. Actually, a solution to (\ref{Chap2AdimensionalIntegralEquation}) is only a function of the argument $x=p/\alpha$. As a consequence, the required function $\Psi(p)$ solving (\ref{Chap2DimensionalIntegralEquation}) is then simply obtained by the formula $\Psi(p)=\psi(p/\alpha)$. 
Contrary to the case of \cal{QED}$_{3+1}$, the rescaled solution $\Psi(\lambda p)$ with $\lambda>0$ does not obey the same equation as $\Psi(p)$, i.e. equation (\ref{Chap2DimensionalIntegralEquation}). It is only a solution in a theory where $e^2$ is changed to $e^{2}/\lambda$. Therefore $\Psi(\lambda p)$ is not a stationary point of the energy  (\ref{Chap2EnergyDensity}).

In actual fact, nothing guarantees that the physical solution is $\Psi\neq 0$, rather than $\Psi= 0$. However, we could wonder if the condensate is energetically more favourable compared to empty Fock vacuum.
If a non trivial solution to (\ref{Chap2DimensionalIntegralEquation}) exists, its energy density will be negative and hence lower than the energy density of the Fock vacuum $|0\rangle$, as we shall briefly show. The substitution of the integral equation (\ref{Chap2IntegralEquation}) in the formula for the energy density $E=\langle\Psi|\hat{H}_{F}|\Psi\rangle/(2\pi)^{2}\delta^{(2)}_{(p)}(0)$ given by (\ref{Chap2EnergyDensity}), where the infinite constant (\ref{Chap2InfiniteConstant}) has been subtracted out, gives the negative value
\begin{eqnarray}
E=\frac{e^{2}}{2}\int \frac{\rmd^{2}k^{i}\rmd^{2}\ell^{i}}{(2\pi)^{4}}\frac{-1}{(\vec{\ell}-\vec{k})^{2}}\frac{\Psi(k)\Psi(\ell)}{(1+\Psi(k)^{2})(1+\Psi(\ell)^{2})}(\Psi(k)\hat{k}-\Psi(\ell)\hat{\ell})^{2}.\nonumber
\end{eqnarray}
Since this energy density is less than the energy density of the Fock vacuum, we may expect that the Fock vacuum will be unstable to decay into the condensate state.
\subsection{Numerical solution}
\begin{figure}[!ht]
\centering
\includegraphics[scale=0.8]{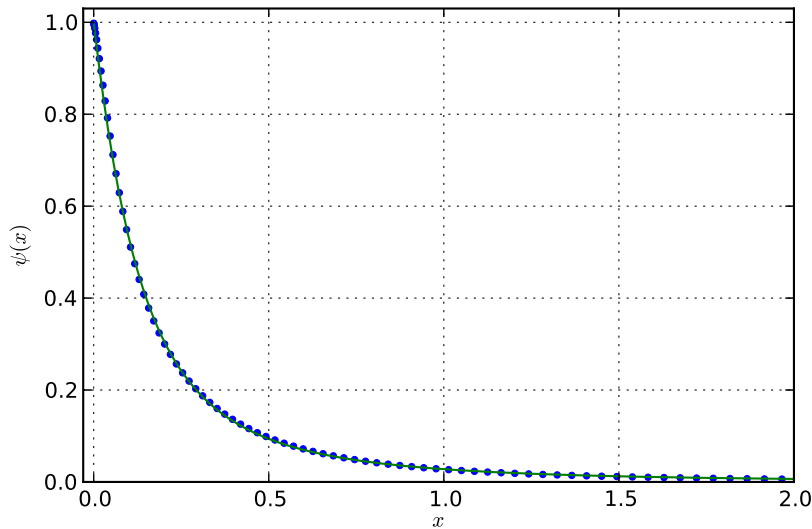}
\caption{The figure compares the trial function (continuous line) with the value of the integral on the {\it rhs} of (\ref{Chap2AdimensionalIntegralEquation}) (dots), as a function of  $x=p/\alpha$.\label{Chap2FigurePsi}} 
\end{figure}

A numerical iteration procedure can produce an approximate solution to the integral equation (\ref{Chap2AdimensionalIntegralEquation}), written in the form
\begin{eqnarray}
\psi(x)=O[\psi](x)
\end{eqnarray}
where $O$ denotes the nonlinear integral operator which can be read from (\ref{Chap2AdimensionalIntegralEquation}).
The numerical recipy consists in finding the best trial function to solve the integral equation. An analytic formula for the wave function depending on a series of parameters was guessed and the values of the parameters were determined by an optimization procedure minimizing the squared difference between the trial function and the {\it rhs} of (\ref{Chap2AdimensionalIntegralEquation}) evaluated on a lattice of points. The approximate solution is illustrated in Fig. \ref{Chap2FigurePsi}.

\subsection{Spontaneous parity violation}
In the literature, reliable arguments support the absence of parity violation (or a parity anomaly) at the perturbative level \cite{RaoParityAnomalies,DelCima}, given massless fermions in the bare Lagrangian. Nonetheless, it is not unexpected that non-perturbative effects may dynamically break this discrete symmetry, as claimed already in \cite{Hoshino:1989ct}. Incidentally, the question of spontaneous parity violation has also been studied in the context of multi-flavour \cal{QED}$_{3}$ (see for example \cite{Lo:2010fm}).

The expectation value of the parity odd operator $:\bar{\chi}(x)\chi(x):$ is vanishing in the Fock vacuum $|0\rangle$. However, the same is not true for the pair condensate $|\Psi\rangle$. The expectation value in the condensate may be calculated with the help of (\ref{MatrixElement1}) leading to
\begin{eqnarray}
&\langle  \Psi|:\bar{\chi}(0,\vec{x})\chi(0,\vec{x}):|\Psi\rangle &= -\int \frac{\rmd^{2}p^{i}}{(2\pi)^{2}}\frac{2\Psi(p)}{1+\Psi(p)^{2}}\\
&&=-(\frac{e^{2}}{4\pi})^{2}\int_{0}^{+\infty}\frac{\rmd y}{\pi}\frac{y\psi(y)}{1+\psi(y)^{2}}.
\end{eqnarray}
A quadrature using the numerical approximation for the condensate wave function gives the following result for the order parameter
\begin{eqnarray}
\langle \Psi|\bar{\chi}\chi|\Psi\rangle \approx -3.2 \cdot  10^{-2}\Big(\frac{e^{2}}{4\pi}\Big)^{2}.
\end{eqnarray}
Hence we conclude that the vacuum $|\Psi\rangle$, which is energetically more favoured, violates parity, as a straightforward consequence of the definition of the coherent state.
Incidentally, the reader will notice that because $\langle 0|:\bar{\chi}(0,\vec{x})\chi(0,\vec{x}):|0\rangle=0$, we have 
\begin{eqnarray}
\langle \Psi|\bar{\chi}(0,\vec{x})\chi(0,\vec{x})|\Psi\rangle=\langle \Psi|:\bar{\chi}(0,\vec{x})\chi(0,\vec{x}):|\Psi\rangle.
\end{eqnarray}

\section{Definition of the Hamilton operator of the quasi-particles\label{QuasiParticleHamiltonian}}
The full quantum Hamiltonian is not yet thoroughly specified. Actually, it may be written completely in terms of the Bogoliubov operators, and should be defined so that its matrix elements are finite. Given that the wave function $\Psi(p)$ is real, one finds the exact result
\begin{eqnarray}
&\hat{H}_{F}=\int \rmd ^{2}p^{i} \ \omega(p)[B^{\dagger}(\vec{p})B(\vec{p})+D^{\dagger}(-\vec{p})D(-\vec{p})]+\langle \Psi|\hat{H}_{F}|\Psi\rangle+:\hat{H}_{C}:_{\Psi}+\nonumber\\
&+2\int \frac{\rmd ^{2}p}{1+\Psi(p)^2}\Big\{p\Psi(p)-\frac{e^{2}}{8\pi^{2}}\int\frac{ \rmd^{2}q^{i} }{(\vec{q}-\vec{p})^{2}}\Big[(1-\Psi(p)^{2})\frac{\Psi(q)}{1+\Psi(q)^{2}}+\nonumber\\
&\quad\quad\qquad \quad +\hat{q}.\hat{p}\ \Psi(p)\frac{\Psi(q)^{2}-1}{\Psi(q)^{2}+1}\Big]\Big\}[B^{\dagger}(\vec{p})D^{\dagger}(-\vec{p})+D(-\vec{p})B(\vec{p})],\label{Chap2QHamiltonian1}
\end{eqnarray}
where the dispersion relation for the quasi-particles is given by the expression
\begin{eqnarray}
&\omega(p)=p\frac{1-\Psi(p)^{2}}{1+\Psi(p)^{2}}& \nonumber\\
&+\frac{e^{2}}{2}\mathcal{P}\int \frac{\rmd ^{2}q}{(2\pi)^{2}}\frac{4\Psi(p)\Psi(q)+\hat{p}.\hat{q}(1+\Psi(q)^{2}\Psi(p)^{2}-\Psi(p)^{2}-\Psi(q)^{2})}{(\vec{p}-\vec{q})^{2}(1+\Psi(p)^{2})(1+\Psi(q)^{2})}& .\label{Chap2DispersionRelationInfrared}
\end{eqnarray}
The Coulomb interaction Hamiltonian $:\hat{H}_{C}:_{\Psi}$ may not be put into a simple form. Examining the quantum Hamiltonian more closely, the new bilinear terms in the first line of (\ref{Chap2QHamiltonian1}) result from the reorganization of the whole Hamiltonian given in (\ref{Chap2QHamiltonian}) and (\ref{Chap2QHamiltonianCoulomb}) as a sum of terms in the normal ordered form associated to the condensate, $:\ :_{\Psi}$.
Hence, this reorganization generates diagonal terms multiplied by a new dispersion relation $\omega(|\vec{p}|)$ as well as off-diagonal terms.
As a consequence of the integral equation (\ref{Chap2IntegralEquation}), the off-diagonal terms in the expression for $\hat{H}_{F}$, which are of the type $B^{\dagger}(\vec{p})D^{\dagger}(-\vec{p})$ or $D(-\vec{p})B(\vec{p})$, vanish so that we can interpret the function $\omega(p)$ as the energy of an excitation of one ``constituent'' fermion or ``quasi-particle'', $B^{\dagger}(\vec{p})|\Psi\rangle$ or $D^{\dagger}(-\vec{p})|\Psi\rangle$. The energy dispersion relation of a quasi-particle may be rewritten as the sum of a finite and a gauge dependent contribution (the latter being potentially divergent),
\begin{eqnarray}
&\omega(p)&=p\frac{1-\Psi(p)^{2}}{1+\Psi(p)^{2}}\label{Chap2DispersionLine1}\\
&&+\frac{e^{2}}{2}\int \frac{\rmd ^{2}q^{i}}{(2\pi)^{2}}\frac{4\Psi(p)\Psi(q)-2\hat{p}.\hat{q}(\Psi(p)^{2}+\Psi(q)^{2})}{(\vec{p}-\vec{q})^{2}(1+\Psi(p)^{2})(1+\Psi(q)^{2})}\label{Chap2DispersionLine2}\\
&&+\frac{e^{2}}{2}\mathcal{P}\int \frac{\rmd ^{2}q^{i}}{(2\pi)^{2}}\frac{\hat{p}.\hat{q}}{(\vec{p}-\vec{q})^{2}}\label{Chap2DispersionLine3}
\end{eqnarray}
where the contribution (\ref{Chap2DispersionLine1}) is the ``corrected'' linear dispersion relation of a relativistic fermion with an asymptotic linear behaviour at large momenta, while the term  (\ref{Chap2DispersionLine2}) is a pure effect of the presence of the pair condensate.  Actually, the integral (\ref{Chap2DispersionLine2}) is convergent whenever $p>0$, but diverges for $p=0$. \\
In order to unravel the low momentum behaviour of the dispersion relation, a closer analysis of the behaviour of this integral at $p\to 0$ is required. We decide to perform the angular integration and to use a limited series expansion of the solution for the wave function
\begin{eqnarray}
\Psi(k)=1+\Psi'(0)k+\dots
\end{eqnarray}
where $k=q$ or $k=p$ is in the interval $[0,\eta]$, while $\eta$ is estimated by looking at the numerical solution. To be more specific, we find that the linear approximation is valid when $\eta\approx 0.1 \alpha=0.1 (e^{2}/4\pi)$. In order to study the singular contribution as $p\to 0$, we limit the radial integral in  (\ref{Chap2DispersionLine2}) to the range $|\vec{q}|\in[0,\eta]$. The integration can then be performed and the result shows that the divergent contribution of (\ref{Chap2DispersionLine2}) behaves like 
\begin{eqnarray}
\frac{e^{2}}{4\pi}\Big\{1-\ln2+\ln(\frac{p+\eta}{2p})-\frac{\Psi'(0)^{2}}{4}\eta^{2}+\dots\label{Chap2NonPerturbativeSmallp}\Big\}
\end{eqnarray}
 where the dots mean that we neglected terms vanishing in the limit $p\to 0$. The result of this approximation is that in the small $p$ region the leading (divergent) behaviour of  (\ref{Chap2DispersionLine2}) is 
\begin{eqnarray}
\frac{e^{2}}{2}\int \frac{\rmd ^{2}q^{i}}{(2\pi)^{2}}\frac{4\Psi(p)\Psi(q)-2\hat{p}.\hat{q}(\Psi(p)^{2}+\Psi(q)^{2})}{(\vec{p}-\vec{q})^{2}(1+\Psi(p)^{2})(1+\Psi(q)^{2})}\sim_{p\ll \eta} \frac{e^{2}}{4\pi}\ln(\frac{\eta}{2p}).
\end{eqnarray}
Therefore, the conclusion is that the influence of the condensate induces a divergent contribution to the energy dispersion relation in the infrared region.

In order to understand the origin of the term (\ref{Chap2DispersionLine3}), it may be instructive to come back to the ordering prescription chosen for the definition of the Coulomb Hamiltonian of the form $-\frac{e^{2}}{2}\rho \Delta^{-1} \rho$ with $\rho=\chi^{\dagger}\chi$. In the quantum Hamiltonian, each charge density was ordered separately, i.e. we chose to define the Hamitonian as follows $:\rho: \Delta^{-1} :\rho:$ which had the advantage to remove the gauge dependence.

As may be observed from (\ref{Chap2OrderingHC}), the difference between this prescription and the choice to order the whole expression $ :\rho \Delta^{-1} \rho:$ is the sum of a constant term (full contraction) and two bilinear terms. Considering only the two bilinear terms in (\ref{Chap2OrderingHC}), a straightforward calculation gives 
\begin{eqnarray}
&-\frac{e^{2}}{2}\int \rmd^{2}x^{i}\rmd^{2}y^{i}&\Big[:\chi^{\dagger}_{\alpha}(\vec{x})\chi_{\alpha}\begC1{(\vec{x})}\conC{G(\vec{x},\vec{y})\chi^{\dagger}_{\beta}}\endC1{(\vec{y})}\chi_{\beta}(\vec{y}):\nonumber\\
&&+:\chi_{\alpha}(\vec{x})\chi^{\dagger}_{\alpha}\begC1{(\vec{x})}\conC{G(\vec{x},\vec{y})\chi_{\beta}}\endC1{(\vec{y})}\chi^{\dagger}_{\beta}(\vec{y}):\Big]\nonumber\\
&=\frac{e^{2}}{2}\int \rmd^{2}p^{i}\rmd^{2}q^{i}&\frac{\hat{p}.\hat{q}}{(\vec{p}-\vec{q})^{2}}\Big[b^{\dagger}(\vec{p})b(\vec{p})+d^{\dagger}(-\vec{p})d(-\vec{p})\Big],\label{Chap2DivergentBilinear}
\end{eqnarray}
which is exactly the extra contribution in $b^{\dagger}(\vec{p})b(\vec{p})+d^{\dagger}(-\vec{p})d(-\vec{p})$ remaining when $\Psi(p)$ is sent to zero in the expression for (\ref{Chap2QHamiltonian1}). This means that the term in (\ref{Chap2DispersionLine3}) is only a consequence of the choice of ordering in the definition of $\hat{H}_{C}$ in (\ref{Chap2QHamiltonianCoulomb}) and hence is not caused by the presence of the condensate. In fact, the operator (\ref{Chap2DivergentBilinear}) has to be understood as the ``finite part'' and is proportional to the diagram

\begin{center}
\fcolorbox{white}{white}{
  \begin{picture}(100,40) (275,-151)
    \SetWidth{1.0}
    \SetColor{Black}
    \Line[arrow,arrowpos=0.5,arrowlength=5,arrowwidth=2,arrowinset=0.2](271,-148)(352,-148)
    \PhotonArc[clock](312,-153.632)(24.652,166.795,13.205){3.5}{8}
   \Vertex(336.652,-148){2}
   \Vertex(287.348,-148){2}
  \end{picture}
}
\end{center}
where the wavy line is associated to the instantaneous ``photon'' propagator (similar to the Coulomb gauge photon) as explained in \ref{Self-EnergySigma}.
Incidentally, we may now notice that an infrared divergence appears if we made the choice of the naive Green function as in (\ref{Chap2NaiveGreenFunction}).
However the Fourier transform of the Green function is actually given by the finite part
\begin{eqnarray}
\frac{e^{2}}{2}\mathcal{P}\int \frac{\rmd ^{2}q^{i}}{(2\pi)^{2}}\frac{\hat{p}.\hat{q}}{(\vec{p}-\vec{q})^{2}}
\end{eqnarray}
as explained before. We find
\begin{eqnarray}
&\frac{e^{2}}{2}\int_{|\vec{p}-\vec{q}|>\mu} \frac{\rmd ^{2}q^{i}}{(2\pi)^{2}}\frac{\hat{p}.\hat{q}}{(\vec{p}-\vec{q})^{2}}+\frac{e^{2}}{2}\int_{|\vec{p}-\vec{q}|<\mu} \frac{\rmd ^{2}q^{i}}{(2\pi)^{2}}\frac{\hat{p}.\hat{q}-1}{(\vec{p}-\vec{q})^{2}}\nonumber\\
&=\frac{e^{2}}{4\pi}\Big[\ln\frac{2p}{\mu}+\ln2-1\Big]\label{Chap2RegularizedSmallp}
\end{eqnarray}
where $p=|\vec{p}|$ and $q=|\vec{q}|$. The details of the calculation leading to (\ref{Chap2RegularizedSmallp}) are given in \ref{Self-EnergySigma}.
The scale $\mu$ is related to the scale present in the logarithm in the Coulomb Green function in $x$-space.  The relation between the scales is given by (\ref{Chap2GreenEpsilon}).

\begin{figure}[!ht]
\centering
\includegraphics[scale=0.8]{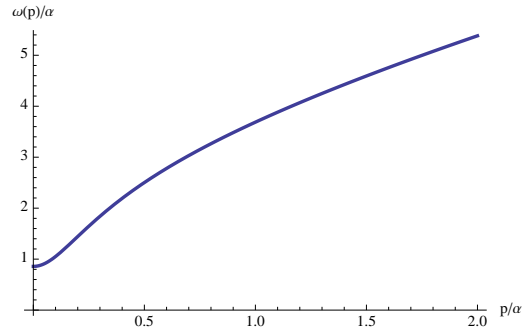}
\caption{The ``renormalized'' dispersion relation in unit of $\alpha=e^{2}/4\pi$, where we chose $\mu=0.1 e^{2}/4\pi$.\label{Chap2FigureDispersion}} 
\end{figure}

Without further ado, we may now study the small $p$ behaviour of the dispersion relation, by summing (\ref{Chap2NonPerturbativeSmallp}) and (\ref{Chap2RegularizedSmallp}), to note that the divergent contributions coming from the logarithms cancel each other. This is confirmed by the numerical evaluation of the dispersion relation as plotted in Fig. \ref{Chap2FigureDispersion}.

\begin{figure}[!ht]
\centering
\includegraphics[scale=0.8]{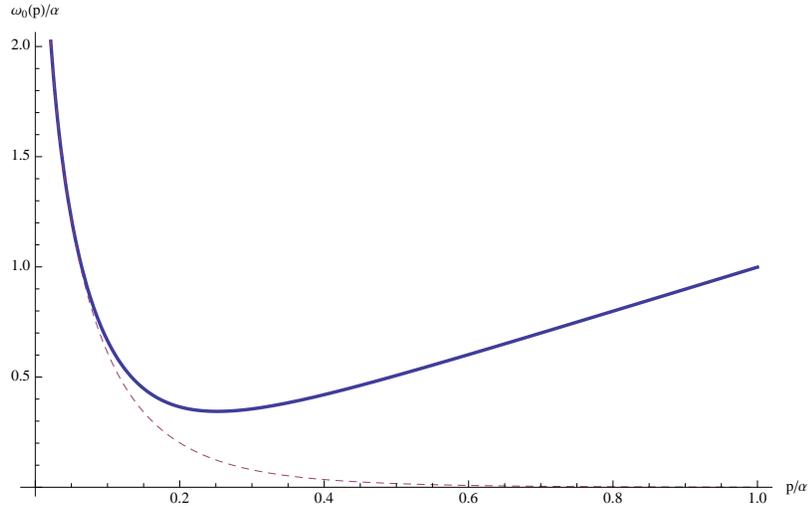}
\caption{The dispersion relation $\omega_{0}$ (thick line) in units of $\alpha=e^{2}/4\pi$. The dashed line represents the contribution of the term (\ref{Chap2DispersionLine2}) to the dispersion relation. \label{Chap2DispersionRelationConfinement}} 
\end{figure}

To disentangle this situation, we may decide to separate the contribution coming from the condensate and the one originating from the self-energy as follows
\begin{eqnarray}
\omega(|\vec{p}|)=\omega_{0}(|\vec{p}|)+\sigma(|\vec{p}|),\label{Chap2DispersionSelfEnergySplit1}
\end{eqnarray}
with
\begin{eqnarray}
\sigma(|\vec{p}|)=\frac{e^{2}}{2}\mathcal{P}\int \frac{\rmd ^{2}q^{i}}{(2\pi)^{2}}\frac{\hat{p}.\hat{q}}{(\vec{p}-\vec{q})^{2}}.\label{Chap2sigmaselfenergy}
\end{eqnarray}
The contribution from the condensate causes a low momentum divergence of the energy as we explained before. This behaviour is illustrated in the Figure (\ref{Chap2DispersionRelationConfinement}), where the rise of the energy as $p\to 0$ is viewed as the signature of the confinement of charges. This will be made clear when we will study the energy of a state made of a pair of opposite charges.

In conclusion, we found that the contribution to the dispersion relation coming from the interaction with the condensate and the contribution coming from the self-energy had the exact opposite behaviour at small momentum. Hence a complete screening of the low momentum divergence is observed. The result is that $\omega(0)$ takes a finite value, which depends on the scale $\mu$. This is not unexpected since the self-energy takes into account the interaction of the particle with its own Coulomb potential which is $\mu$-dependent. By the way, a similar screening of divergencies is described in \cite{Cornwall:1980zw}, based on a different treatment.

The dependence on $\mu$ is the fingerprint of the confining electrostatic potential, and is justified in the expression for the energy of a single charged particle because, by itself a state composed of a single charged particle is not gauge invariant. Nevertheless, in complete analogy with the classical situation, we will show hereafter that the mean energy of a particle/anti-particle pair is independent of $\mu$ and it is neither UV divergent, nor IR divergent. In Section \ref{SectionGreenFunctions}, we will show that we can understand $\omega_{0}(p)$ as the energy at a pole of the fermion propagator dressed by the Coulomb interaction.
\section{Residual Coulomb interactions \label{SectionEnergyPair}}
As a matter of fact, the energy of a state composed of a single charged particle depends on the scale $\mu$ present in the Coulomb Green function. The reason for this observation is that such a state is not physical. On the contrary, a charge neutral state is physical and should have a gauge invariant energy.
The goal of this section is to show that a bound state of the form
\begin{eqnarray}
|f\rangle=\int \rmd^{2}k^{i}f(|\vec{k}|)B^{\dagger}(\vec{k})D^{\dagger}(-\vec{k})|\Psi\rangle,
\end{eqnarray}
has a finite Coulomb energy. The pair state can be interpreted as a positronium, at rest in the ``center of mass'' frame. 
As a perspective, once the value for the bound state energy is established, a ``Schr\"odinger'' equation can be derived from the variation
\begin{eqnarray}
\frac{\delta}{\delta f^{*}(k)}\frac{\langle f|\hat{H}_{0}|f\rangle}{\langle f|f\rangle}=0,\label{Chap2BoundStateEqn}
\end{eqnarray}
where the magnetic mode sector is ignored and considering a simplified Hamiltonian
\begin{eqnarray}
\hat{H}_{0}=\hat{H}_{K}+:\hat{H}_{C}:_{\Psi},
\end{eqnarray}
with
\begin{eqnarray}
\hat{H}_{K}=\int \rmd^{2}p^{i}\omega(|\vec{p}|)[B^{\dagger}(\vec{p})B(\vec{p})+D^{\dagger}(-\vec{p})D(-\vec{p})].
\end{eqnarray}
The solution of this integral equation would provide the wave function $f(|\vec{p}|)$ and the energy of the lowest excitation of the bound state. For instance, the numerical procedure could involve a Gauss-Laguerre quadrature method, which leads to a non-trivial problem, even in absence of a pair condensate. However we leave this possibility for future work.

In order to evaluate the  energy of the bound state and before calculating the Coulomb interaction energy, a first trivial result is
\begin{eqnarray}
\frac{\langle f|\hat{H}_{K}|f\rangle}{(2\pi)^{2}\delta^{(2)}_{(p)}(0)}=\int \frac{\rmd^{2}k^{i}}{(2\pi)^{2}}2\omega(|\vec{k}|)|f(k)|^{2}
\end{eqnarray}
where we decide to explicitly single out two terms in the dispersion relation
\begin{eqnarray}
\omega(|\vec{k}|)=\omega_{0}(|\vec{k}|)+\frac{e^{2}}{2}\mathcal{P}\int \frac{\rmd ^{2}q^{i}}{(2\pi)^{2}}\frac{\hat{k}.\hat{q}}{(\vec{k}-\vec{q})^{2}},\label{Chap2DispersionSelfEnergySplit}
\end{eqnarray}
which corresponds to a separation of the $\mu$-dependent parts contributing to the dispersion relation.

\subsection{Calculation of the residual Coulomb interactions}
In order to compute the residual Coulomb interactions given by $:\hat{H}_{C}:_{\Psi}$, we will provide the details essential to obtain the necessary expressions.
A little algebra shows that
\begin{eqnarray}
&:\chi^{\dagger}(\vec{x})\chi(\vec{x}):=\int \frac{\rmd^{2}k^{i}\rmd^{2}\ell^{i}}{(2\pi)^{2}\sqrt{2|\vec{k}|2|\vec{\ell}|}}e^{\rmi (\vec{\ell}-\vec{k}).\vec{x}}\nonumber\\
&\Big\{M_{1}(\vec{k},\vec{\ell})B^{\dagger}(\vec{k})B(\vec{\ell})-M_{2}(\vec{k},\vec{\ell})D^{\dagger}(-\vec{\ell})D(-\vec{k})\nonumber\\
 &\quad\qquad+M_{3}(\vec{k},\vec{\ell})B^{\dagger}(\vec{k})D^{\dagger}(-\vec{\ell})+M_{4}(\vec{k},\vec{\ell})D(-\vec{k})B(\vec{\ell})\Big\}
\end{eqnarray}
where we have defined the following functions of the wave function of the condensate
\begin{eqnarray*}
&M_{1}(\vec{k},\vec{\ell})=&u^{\dagger}(\vec{k})u(\vec{\ell})[\alpha(k)\alpha(\ell)+\beta(l)\beta(k)]\\
&&-u^{\dagger}(\vec{k})u(-\vec{\ell})[\alpha(k)\beta(l)+\alpha(\ell)\beta(k)]\\
&M_{2}(\vec{k},\vec{\ell})=&u^{\dagger}(\vec{k})u(\vec{\ell})[\alpha(k)\alpha(\ell)+\beta(\ell)\beta(k)]\\
&&+u^{\dagger}(\vec{k})u(-\vec{\ell})[\alpha(\ell)\beta(k)+\alpha(k)\beta(\ell)]\\
&M_{3}(\vec{k},\vec{\ell})=&u^{\dagger}(\vec{k})u(\vec{\ell})[\alpha(k)\beta(\ell)-\alpha(\ell)\beta(k)]\\
&&+u^{\dagger}(\vec{k})u(-\vec{\ell})(\alpha(k)\alpha(\ell)-\beta(k)\beta(\ell)]\\
&M_{4}(\vec{k},\vec{\ell})=&u^{\dagger}(\vec{k})u(\vec{\ell})[\alpha(\ell)\beta(k)-\alpha(k)\beta(\ell)]\\
&&+u^{\dagger}(\vec{k})u(-\vec{\ell})[\alpha(k)\alpha(\ell)-\beta(k)\beta(\ell)].
\end{eqnarray*}
For the sake of completeness, we also give the following results
\begin{eqnarray}
&u^{\dagger}(\vec{k})u(\vec{\ell})=\sqrt{|\vec{k}|.|\vec{\ell}|}(1+\hat{k}.\hat{\ell}+\rmi \hat{\ell}\times\hat{k}),\\
&u^{\dagger}(\vec{k})u(-\vec{\ell})=\sqrt{|\vec{k}|.|\vec{\ell}|}(1-\hat{k}.\hat{\ell}-\rmi \hat{\ell}\times\hat{k}).
\end{eqnarray}
We shall now consider the interactions involving only one pair. Among all the possible Coulomb interactions, we find that the only terms contributing to (\ref{Chap2BoundStateEqn}) are
\begin{eqnarray}
&-\frac{e^{2}}{2}\int\rmd^{2}x^{i}\rmd^{2}y^{i}\Big[:(:\chi^{\dagger}(\vec{x})\chi(\vec{x}):G(\vec{x},\vec{y}):\chi^{\dagger}(\vec{y})\chi(\vec{y}):):_{\Psi}\Big]_{1P}\nonumber\\
&=-\frac{e^{2}}{2}\mathcal{P}\int \frac{\rmd^{2}\ell^{i}\rmd^{2}k^{i}\rmd^{2}p^{i}}{(2\pi)^{2}}\frac{-1}{|\vec{p}|^{2}}\frac{2}{\sqrt{2|\vec{\ell}|2|\vec{k}|2|\vec{\ell}+\vec{p}|2|\vec{k}-\vec{p}|}}\quad\quad\nonumber\\
&\Big\{-M_{1}(\vec{k},\vec{k}-\vec{p})M_{2}(\vec{\ell},\vec{\ell}+\vec{p})\times\quad\quad\quad\nonumber\\
&\quad\quad\times B^{\dagger}(\vec{k})D^{\dagger}(-(\vec{\ell}+\vec{p}))D(-\vec{\ell})B(\vec{k}-\vec{p})\label{Chap21PairHamiltonian1}\\
&+M_{3}(\vec{k},\vec{k}-\vec{p})M_{4}(\vec{\ell},\vec{\ell}+\vec{p})\times\quad\quad\quad\nonumber\\
&\quad\quad\times B^{\dagger}(\vec{k})D^{\dagger}(-(\vec{k}-\vec{p}))D(-\vec{\ell})B(\vec{\ell}+\vec{p})\Big\}.\label{Chap21PairHamiltonian2}
\end{eqnarray}
The following useful matrix element of the residual Coulomb Hamiltonian can be separated in two terms, corresponding to the first and second terms, respectively (\ref{Chap21PairHamiltonian1}) and (\ref{Chap21PairHamiltonian2}),
\begin{eqnarray}
\frac{\langle f|(:\hat{H}_{C}:_{\Psi})_{1P}|f\rangle}{(2\pi)^{2}\delta^{(2)}_{(p)}(0)}=T_{1}+T_{2},
\end{eqnarray}
where $T_{1}$ corresponds to the one Coulomb photon exchange inside the pair

\begin{center}
\fcolorbox{white}{white}{
  \begin{picture}(66,38) (95,-77)
    \SetWidth{1.0}
    \SetColor{Black}
    \Line[arrow,arrowpos=0.5,arrowlength=5,arrowwidth=2,arrowinset=0.2](96,-42)(128,-42)
    \Line[arrow,arrowpos=0.5,arrowlength=5,arrowwidth=2,arrowinset=0.2](128,-42)(160,-42)
    \Line[arrow,arrowpos=0.5,arrowlength=5,arrowwidth=2,arrowinset=0.2](128,-74)(160,-74)
    \Line[arrow,arrowpos=0.5,arrowlength=5,arrowwidth=2,arrowinset=0.2](96,-74)(128,-74)
    \Photon(128,-42)(128,-74){2.5}{4}
    \Vertex(128,-42){2}
    \Vertex(128,-74){2}
  \end{picture}
}
\end{center}
and where $T_{2}$ is associated to the annihilation 

\begin{center}
\fcolorbox{white}{white}{
  \begin{picture}(67,35) (95,-62)
    \SetWidth{1.0}
    \SetColor{Black}
    \Line[arrow,arrowpos=0.5,arrowlength=5,arrowwidth=2,arrowinset=0.2](96,-29)(112,-45)
    \Line[arrow,arrowpos=0.5,arrowlength=5,arrowwidth=2,arrowinset=0.2](96,-61)(112,-45)
    \Photon(112,-45)(145,-45){2.5}{4}
    \Vertex(112,-45){2}
    \Vertex(145,-45){2}
    \Line[arrow,arrowpos=0.5,arrowlength=5,arrowwidth=2,arrowinset=0.2](145,-45)(161,-29)
    \Line[arrow,arrowpos=0.5,arrowlength=5,arrowwidth=2,arrowinset=0.2](145,-45)(161,-61)
  \end{picture}
}
\end{center}
of the pair into a Coulomb photon. We choose to study only the contribution of $T_{1}$, because $T_{2}$ is independent of the choice of zero of the potential $\mu$. The inclusion of $T_{2}$ in the discussion is nonetheless straightforward.
We find a result with a potential IR divergence at $\vec{k}=\vec{\ell}$, however the integration is considered as the ``finite part'',
\begin{eqnarray}
&T_{1}&=\frac{e^{2}}{2}\mathcal{P}\int \frac{\rmd^{2}\ell^{i}\rmd^{2}k^{i}}{(2\pi)^{4}}\frac{-1}{(\vec{k}-\vec{\ell})^{2}}f^{*}(|\vec{k}|)f(|\vec{\ell}|)\times\nonumber\\
& &\times\Big[1+\hat{\ell}.\hat{k}-2\frac{\psi(\ell)^{2}+\psi(k)^{2}-2\hat{\ell}.\hat{k}\psi(\ell)\psi(k)}{(1+\psi(\ell)^{2})(1+\psi(k)^{2})}\Big].\label{Chap2T1}
\end{eqnarray}
In the last equation (\ref{Chap2T1}), we have split the contribution coming from the pair condensate from the one already present in the Fock vacuum. 
The need for the ``finite part'' introduces a $\mu$ dependence in the expression.
One may notice that the contribution coming from the condensate in (\ref{Chap2T1}) vanishes when $\vec{k}=\vec{\ell}$, while the term $1+\hat{\ell}.\hat{k}$ is divergent if we set $\mu=0$. The equation (\ref{Chap2T1}), when evaluated with $\Psi(p)=0$, is completely analogous to the formula found in \cite{Finger:1979yt} which analysed a similar situation in 3+1 dimensions in the so-called Limited Fock Space Approximation. However, in $3+1$ dimensions, no infrared divergence is expected when $\vec{k}=\vec{\ell}$ because the double angular integration makes the singularity integrable. In this case the IR singularity behaves like $\ln|\frac{k+l}{k-l}|$.

Considering the sum of the kinetic and interaction mean energies, we find
\begin{eqnarray}
&E=\frac{\langle f|\hat{H}_{K}|f\rangle}{(2\pi)^{2}\delta^{(2)}_{(p)}(0)}+T_{1}=\int \frac{\rmd^{2}k^{i}}{(2\pi)^{2}}2\omega_{0}(|\vec{k}|)|f(k)|^{2}+\\
&+\frac{e^{2}}{2}\int \frac{\rmd^{2}\ell^{i}\rmd^{2}k^{i}}{(2\pi)^{4}}\frac{-1}{(\vec{k}-\vec{\ell})^{2}}\Big\{f^{*}(|\vec{k}|)f(|\vec{\ell}|)\times\nonumber\\
 &\times\Big[1+\hat{\ell}.\hat{k}-2\frac{\psi(\ell)^{2}+\psi(k)^{2}-2\hat{\ell}.\hat{k}\psi(\ell)\psi(k)}{(1+\psi(\ell)^{2})(1+\psi(k)^{2})}\Big]-2\hat{\ell}.\hat{k}|f(|\vec{k}|)|^{2}\Big\}.\label{Chap2EnergyPositronium}
\end{eqnarray}
where the potentially divergent terms in the second term of (\ref{Chap2DispersionSelfEnergySplit}) and in (\ref{Chap2T1}) have cancelled each other. The result is that the finite part is not needed to render the value of the integral infrared finite. We emphasize once more that the mean energy is now independent of the scale $\mu$. Remarquably, from the contribution of the dispersion relation, only the term $\omega_{0}(k)$, which is pictured in Fig. \ref{Chap2DispersionRelationConfinement}, remains. The divergence at $k\to 0$ of $\omega_{0}(k)$ is a signature of confinement, since it forces the wave function $f(k)$ to vanish at small momentum.
We can symmetrize the last term in (\ref{Chap2EnergyPositronium}) and using the identity,
\begin{eqnarray}
|f(k)|^{2}+|f(\ell)|^{2}=|f(k)-f(\ell)|^{2}+f^{*}(l)f(k)+f^{*}(k)f(\ell)
\end{eqnarray}
we may reformulate the energy of the pair state as
\begin{eqnarray}
&E=\int \frac{\rmd^{2}k^{i}}{(2\pi)^{2}}2\omega_{0}(|\vec{k}|)|f(k)|^{2}+\frac{e^{2}}{2}\int \frac{\rmd^{2}\ell^{i}\rmd^{2}k^{i}}{(2\pi)^{4}}\frac{-1}{(\vec{k}-\vec{\ell})^{2}}\Big\{f^{*}(k)f(\ell)\times\nonumber\\
&\times\Big[1-\hat{\ell}.\hat{k}-2\frac{\psi(\ell)^{2}+\psi(k)^{2}-2\hat{\ell}.\hat{k}\psi(\ell)\psi(k)}{(1+\psi(\ell)^{2})(1+\psi(k)^{2})}\Big] -\hat{\ell}.\hat{k}|f(k)-f(\ell)|^{2}\Big\}.\label{Energy1pair}
\end{eqnarray}
This result allows us to conclude that the energy of a state made of a pair of opposite charge particles is indeed independent of the choice of zero of the potential. Therefore we confirm here that the energy of a gauge invariant state is perfectly infrared finite and gauge independent.
By the same token, the examination of the energy of a pair state (\ref{Energy1pair}) confirms the confinement scenario. Since the potential energy between the constituent fermions $2\omega_{0}(|\vec{p}|)$ is divergent at $\vec{p}=\vec{0}$, a likely and reasonable assumption is that  the wave function of the pair, $f(|\vec{p}|)$, vanishes in order to solve the bound state equation
\begin{eqnarray}
&(2\omega_{0}(k)-E)f(k)+\frac{e^{2}}{2}\int \frac{\rmd^{2}\ell^{i}}{(2\pi)^{2}}\frac{-1}{(\vec{k}-\vec{\ell})^{2}}\times\nonumber\\
 &\times\Big\{f(\ell)\Big[1-\hat{\ell}.\hat{k}-2\frac{\psi(\ell)^{2}+\psi(k)^{2}-2\hat{\ell}.\hat{k}\psi(l)\psi(k)}{(1+\psi(\ell)^{2})(1+\psi(k)^{2})}\Big]\nonumber\\
 &-\hat{\ell}.\hat{k}[f(k)-f(\ell)]\Big\}=0,
\end{eqnarray} 
obtained from (\ref{Chap2BoundStateEqn}). The study of the energy levels of this bound state necessitates the numerical solution of this eigenvalue integral equation. Beforehand, the interaction with the magnetic mode should be probably included in the variational principle in order to get a more consistent approximation. The numerical resolution of this equation is left open for future work.

\section{Green function interpretation \label{SectionGreenFunctions}}
\subsection{Schwinger-Dyson equation \label{Schwinger_Dyson}}
The Hamilton formalism has made clear that a variational procedure was an appropriate way to obtain the structure of the fermionic vacuum.
As a complementary point of view on the condensation mechanism, we may understand the integral equation (\ref{Chap2IntegralEquation}) as a (truncated) Schwinger-Dyson equation \cite{GovaertsThesis}. More precisely, the idea is to choose an ansatz for a $p$-space propagator, and to identify the relationship between a Schwinger-Dyson equation and the integral equation for the wave function.
To do so, let us introduce the following formula for the fermion propagator in the condensate
\begin{eqnarray}
S^{(3)}(p^{0},\vec{p})=\frac{\rmi}{\slashed{p}-\Sigma(p)+\rmi\epsilon}
\end{eqnarray}
with the parametrization $\Sigma(p)=|\vec{p}|A(p)+\vec{p}.\vec{\gamma}B(p)$. The functions $A(p)=A(|\vec{p}|)$ and $B(p)=B(|\vec{p}|)$ depend only on the modulus of $\vec{p}$. When we substitute the parametrization in the Feynman propagator, we obtain
\begin{eqnarray}
S^{(3)}(p^{0},\vec{p})=\rmi\frac{p^{0}\gamma^{0}-\vec{p}.\vec{\gamma}(1+B(p))+|\vec{p}|A(p)}{(p^{0})^{2}-|\vec{p}|^{2}(A^{2}(p)+(1+B(p))^{2})+\rmi\epsilon}.\label{Chap2PropagatorCondensate}
\end{eqnarray}
As a consequence, the calculation of the equal time propagator in the condensate
\begin{eqnarray}
S(\vec{p})=\int\frac{\rmd p^{0}}{2\pi}S^{(3)}(p^{0},\vec{p})\label{EquaTimePropagator}
\end{eqnarray}
allows one to the obtain a relation between the wave function of the condensate and the functions $A(p)$ and $B(p)$. More precisely, we find the relation between the wave function of the condensate and the ansatz functions
\begin{eqnarray}
\frac{1}{2}\frac{A(p)+\hat{p}.\vec{\gamma}(1+B(p))}{\sqrt{A^{2}(p)+(1+B(p))^{2}}}=\frac{1}{2}\Big[\frac{2\Psi(p)}{1+\Psi(p)^{2}}+\frac{1-\Psi(p)^{2}}{1+\Psi(p)^{2}}\hat{p}.\vec{\gamma}\Big],
\end{eqnarray}
by integrating (\ref{EquaTimePropagator}) using the parametrization (\ref{Chap2PropagatorCondensate}) and identifying the result with the equal time propagator obtained from (\ref{MatrixElement2}).
This allows to identify 
\begin{eqnarray}
\frac{A(p)}{1+B(p)}=\frac{2\Psi(p)}{1-\Psi(p)^{2}}. 
\end{eqnarray}
In order to fully understand $A(p)$ and $B(p)$ in terms $\Psi(p)$, we need to study the following Schwinger-Dyson equation
\begin{eqnarray}
-\rmi\Sigma(\vec{p})=\int \frac{\rmd^{3}q}{(2\pi)^{3}}\frac{\rmi}{(\vec{p}-\vec{q})^{2}}(\rmi e \gamma^{0})S^{(3)}(q)(\rmi e \gamma^{0}),\label{Chap2SchwingerDysonEquationPropagator}
\end{eqnarray}
which can be pictorially represented as 

\begin{center}
\fcolorbox{white}{white}{
  \begin{picture}(369,70) (175,-40)
    \SetWidth{1.0}
    \SetColor{Black}
    \COval(240,-23)(16,16)(0){Black}{White}
    \Line[arrow,arrowpos=0.5,arrowlength=5,arrowwidth=2,arrowinset=0.2](256,-23)(304,-23)
    \Line[arrow,arrowpos=0.5,arrowlength=5,arrowwidth=2,arrowinset=0.2](176,-23)(224,-23)
    \Text(315,-23)[lb]{\Large{\Black{$=$}}}


    \PhotonArc[clock](416,-23)(48,-180,-360){5.5}{10}
    \Vertex(368,-23){2}
    \Vertex(464,-23){2}
    \GOval(416,-23)(16,16)(0){0.882}
    \Line[arrow,arrowpos=0.8,arrowlength=5,arrowwidth=2,arrowinset=0.2](336,-23)(400,-23)
    \Line[arrow,arrowpos=0.2,arrowlength=5,arrowwidth=2,arrowinset=0.2](432,-23)(496,-23)
  \end{picture}
}
\end{center}
where the Feynman rules for the associated diagrammatic formulation are listed in the  \ref{FeynmanRulesAnnexe}.
The corresponding integral equation may be rewritten
\begin{eqnarray}
pA(p)+\vec{p}.\vec{\gamma}B(p)=\frac{e^{2}}{8\pi^{2}}\int\frac{\rmd^{2}q^{i}}{(\vec{q}-\vec{p})^{2}}\Big[\frac{2\Psi(q)}{1+\Psi(q)^{2}}+\frac{1-\Psi(q)^{2}}{1+\Psi(q)^{2}}\hat{q}.\vec{\gamma}\Big],\label{Chap2SchwingerDyson}
\end{eqnarray}
with $p=|\vec{p}|$ and where the integral should be understood as the Hadamard finite part.
If we multiply (\ref{Chap2SchwingerDyson}) by 
\begin{eqnarray}
N_{1}(\vec{p})=\frac{(1+B(p))+\hat{p}.\vec{\gamma}A(p)}{\sqrt{A^{2}(p)+(1+B(p))^{2}}}=\frac{1-\Psi(p)^{2}}{1+\Psi(p)^{2}}+\hat{p}.\vec{\gamma}\frac{2\Psi(p)}{1+\Psi(p)^{2}},
\end{eqnarray}
and take the trace, we obtain the integral equation (\ref{Chap2IntegralEquation}), as could have been anticipated,
\begin{eqnarray}
4p\Psi(p)=\frac{e^{2}}{2\pi^{2}}\int\frac{ \rmd^{2}q^{i} }{(\vec{q}-\vec{p})^{2}}[(1-\Psi(p)^{2})\frac{\Psi(q)}{1+\Psi(q)^{2}}+\hat{q}.\hat{p}\ \Psi(p)\frac{\Psi(q)^{2}-1}{\Psi(q)^{2}+1}],\nonumber
\end{eqnarray}
 where we used in the calculation the relation
\begin{eqnarray}
\frac{A(p)}{\sqrt{A^{2}(p)+(1+B(p))^{2}}}=\frac{2\Psi(p)}{1+\Psi(p)^{2}}.
\end{eqnarray}
Similarly, we can express the function $A(p)$ in terms of the wave function of the condensate. Taking the trace of (\ref{Chap2SchwingerDyson}) over the spinor indices, we find
\begin{eqnarray}
|\vec{p}|A(p)=\frac{e^{2}}{(2\pi)^{2}}\mathcal{P}\int\rmd^{2}q^{i} \frac{1}{(\vec{p}-\vec{q})^{2}}\frac{\Psi(q)}{1+\Psi(q)^{2}}.
\end{eqnarray}
Finally, we notice that the pole structure of (\ref{Chap2PropagatorCondensate}) provides the energy of the particle excitations: $|\vec{p}|\sqrt{A^{2}(p)+(1+B(p))^{2}}$.
In order to obtain the formula for the dispersion relation, we can add $\vec{p}.\vec{\gamma}$ to (\ref{Chap2SchwingerDyson}) and multiply it by
\begin{eqnarray}
N_{2}(\vec{p})=\frac{A(p)+\hat{p}.\vec{\gamma}(1+B(p))}{\sqrt{A^{2}(p)+(1+B(p))^{2}}}=\frac{2\Psi(p)}{1+\Psi(p)^{2}}-\frac{1-\Psi(p)^{2}}{1+\Psi(p)^{2}}\hat{p}.\vec{\gamma},
\end{eqnarray}
and finally take the trace. The result of this short manipulation gives 
\begin{eqnarray}
&|\vec{p}|\sqrt{A^{2}(p)+(1+B(p))^{2}}=p\frac{1-\Psi(p)^{2}}{1+\Psi(p)^{2}}\nonumber\\
&+\frac{e^{2}}{2}\mathcal{P}\int \frac{\rmd ^{2}q^{i}}{(2\pi)^{2}}\frac{4\Psi(p)\Psi(q)+\hat{p}.\hat{q}(1-\Psi(p)^{2})(1-\Psi(q)^{2})}{(\vec{p}-\vec{q})^{2}(1+\Psi(p)^{2})(1+\Psi(q)^{2})},
\end{eqnarray}
which is exactly the dispersion relation $\omega(p)$ found before in (\ref{Chap2DispersionRelationInfrared}). In conclusion, we obtain that the energy of the quasi-particles created by $B^{\dagger}$ and $D^{\dagger}$ corresponds exactly the energy of the physical pole of the propagator of the fermion field in the pair condensate. Hence this result supports the interpretation obtained before.
In order to complete the analogy, we can reformulate the energy of the condensate in the diagramatic expression

\begin{center}
\fcolorbox{white}{white}{
  \begin{picture}(330,44) (179,-198)
    \SetWidth{1.0}
    \SetColor{Black}
    \Arc[arrow,arrowpos=0.5,arrowlength=5,arrowwidth=2,arrowinset=0.2](250,-180)(15.811,235,595)
    \Arc[arrow,arrowpos=0.5,arrowlength=5,arrowwidth=2,arrowinset=0.2](321,-180)(15.811,235,595)
    \Arc[arrow,arrowpos=0.5,arrowlength=5,arrowwidth=2,arrowinset=0.2](385,-180)(15.811,235,595)
    \Arc[arrow,arrowpos=0.5,arrowlength=5,arrowwidth=2,arrowinset=0.2](447,-180)(15.811,235,595)
    \GOval(371,-180)(6,6)(0){0.882}
    \GOval(265,-180)(6,6)(0){0.882}
    \GOval(399,-180)(6,6)(0){0.882}
    \Photon(385,-164)(385,-195.8){4.5}{3}
    \Vertex(385,-164){1.414}
    \Vertex(385,-195.8){1.414}
    \Vertex(447,-164){1.414}
    \Vertex(447,-195.8){1.414}
    \Photon(447,-164)(447,-195.8){4.5}{3}
    \Vertex(234,-179){1.414}
    \Vertex(305,-181){1.414}
    \Text(176,-183)[lb]{\Black{$E=$}}
    \Text(216,-185)[lb]{\Black{$\vec{p}.\vec{\gamma}$}}
    \Text(278,-185)[lb]{\Black{$-\vec{p}.\vec{\gamma}$}}
    \Text(346,-190)[lb]{\Black{$+ \rmi \Big[$}}
    \Text(474,-190)[lb]{\Black{$ \Big]$}}
    \Text(416,-184)[lb]{\Black{$+$}}
  \end{picture}
}
\end{center}
which can be readily used to obtain the Schwinger-Dyson equation (\ref{Chap2SchwingerDysonEquationPropagator}). The last term in the sum above is a constant divergent bubble diagram that was subtracted from the Hamiltonian when we discussed the value of the energy density of the condensate.

It should be emphasized that the approach developed here does not rely on the dimensional regularization used more or less implicitly in the literature, but on the exact Fourier transform of the $x$-space Coulomb Green function. 

\subsection{Two-point function}
In the previous section, an ansatz technique allowed to obtain a Schwinger-Dyson equation for the fermion propagator. However it is not clear why the propagator obtained in this manner has a gauge dependent pole. In order to explain this issue, it may be more instructive to understand the origin of the ``constituent'' fermion propagator in the condensate from the Fourier transform of a $x$-space correlation function.
Indeed, the relationship between the propagator found above and the approximate (or perturbative) evaluation of a time ordered correlation function is not cristal clear.
The Green function of the equation (\ref{Chap2PropagatorCondensate}), obtained in the $p$-space in the last section, obviously exhibits a pole whose position is gauge dependent. The energy at the pole corresponds to the energy of a single excitation $B^{\dagger}(\vec{p})|\Psi\rangle$ or $D^{\dagger}(-\vec{p})|\Psi\rangle$, in expectation value. From this point of view, it is not a surprise since a charged state is not gauge invariant, however we may raise the question of the mass of these constituent fermions. This puzzle has its origin in the Coulomb interactions. From a perturbative perspective, this feature can be understood as follows. The total quantum Hamiltonian can be splitted in 
\begin{eqnarray}
\hat{H}=\hat{H}_{0}+\hat{H}_{C}^{I} + \hat{H}_{\Phi}+\hat{H}^{I}_{\Phi\chi},\label{TotalQuantumHamiltonian}
\end{eqnarray}
where $ \hat{H}_{\Phi}$ is given by (\ref{FreeMagneticHamiltonian}), and where the Hamiltonian $\hat{H}^{I}_{\Phi\chi}$ is obtained thanks to the Hamiltonian density (\ref{HPhiChi}), while
\begin{eqnarray}
&\hat{H}_{0}&=\int \rmd ^{2}p^{i} \omega_{0}(p)[B^{\dagger}(\vec{p})B(\vec{p})+D^{\dagger}(\vec{p})D(\vec{p})],\label{Chap2gaugesplit1}\\
&\hat{H}_{C}^{I}&=\int \rmd ^{2}p^{i}  \sigma(p)[B^{\dagger}(\vec{p})B(\vec{p})+D^{\dagger}(\vec{p})D(\vec{p})]+:\hat{H}_{C}:_{\Psi},\label{Chap2gaugesplit2}
\end{eqnarray}
where the first term (\ref{Chap2gaugesplit1}) is bilinear and gauge invariant, while the second term (\ref{Chap2gaugesplit2}) is also separately gauge invariant and contains a bilinear and quadrilinear term. The reason of this separation is the ordering prescription taken for the Coulomb Hamiltonian, which insures that the sum of the two gauge dependent terms in (\ref{Chap2gaugesplit2}) is in fact gauge invariant, when acting in the physical state space.

In a perturbative treatment, one should consider $\hat{H}_{0}$ as the ``free'' Hamiltonian, whereas $\hat{H}^{I}_{C}$ and $\hat{H}^{I}_{\Phi\chi}$ as the ``interaction'' Hamiltonians.
In order to define a gauge invariant two-point function in the condensate, we decide to define the interaction picture field
\begin{eqnarray}
\chi_{I}(t,\vec{x})=e^{\rmi\hat{H}_{0}t} \chi(0,\vec{x}) e^{-\rmi\hat{H}_{0}t}.
\end{eqnarray}
The time ordered and gauge invariant two-point function in the condensate can be calculated thanks to 
\begin{eqnarray}
\mathcal{S}(t,\vec{x})=\langle \Psi|\chi_{I}(t,\vec{x})\bar{\chi}_{I}(0,\vec{0})|\Psi\rangle\Theta(t)-\langle \Psi|\bar{\chi}_{I}(0,\vec{0})\chi_{I}(t,\vec{x})|\Psi\rangle\Theta(-t),\nonumber
\end{eqnarray}
where $\Theta(t)$ is the Heaviside step function.
An explicit calculation allows to express the Fourier transform of the two-point function
\begin{eqnarray}
\mathcal{S}(k^{0},\vec{k})=\int\rmd t \ \rmd^{2} x^{i} \ e^{\rmi k^{0}t-\rmi \vec{k}.\vec{x}}\mathcal{S}(t,\vec{x})
\end{eqnarray}
which is represented by a fermion line with a dark blob

\begin{figure}[!ht]
\centering
\subfloat{\label{Chap2m_k}  \includegraphics[scale=0.6]{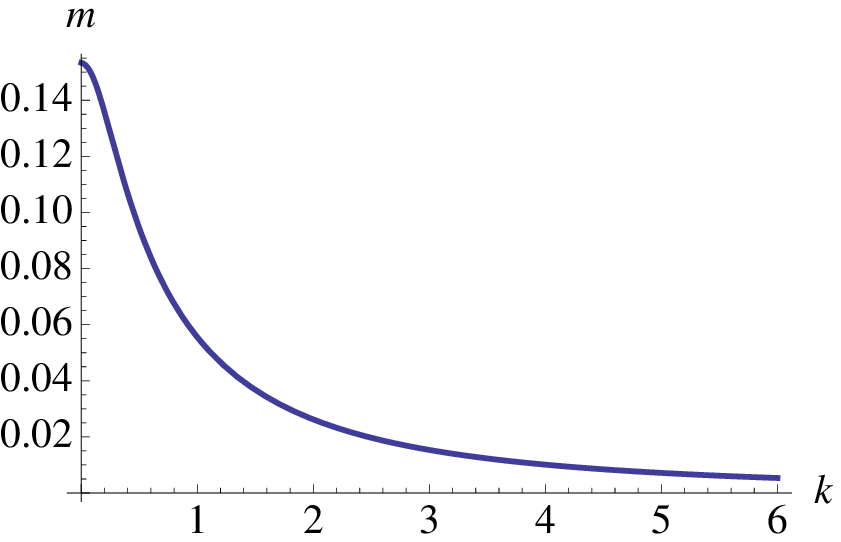} }
  \hspace{8pt}
\subfloat{\label{Chap2Z_k}\includegraphics[scale=0.6]{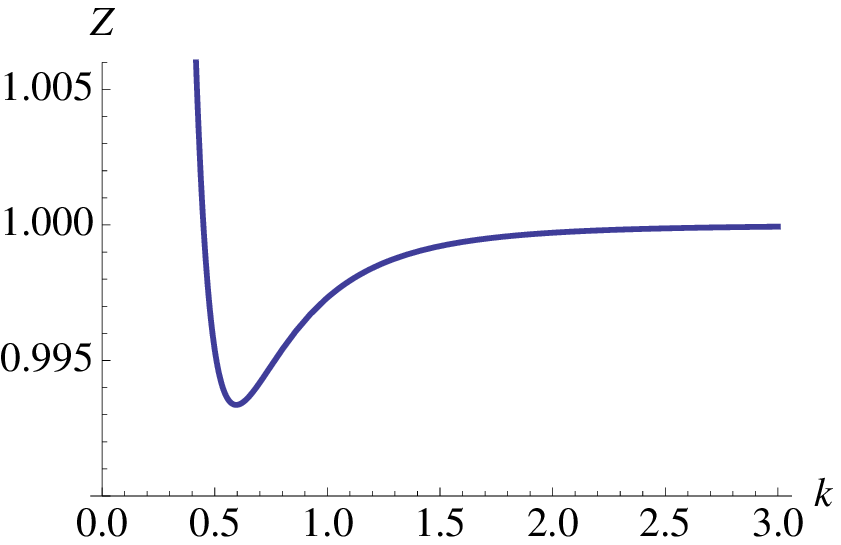}}
\caption{The functions $m(k)$, and $Z(k)$ in units of $e^{2}/4\pi$.}
\end{figure}

\begin{center}
\fcolorbox{white}{white}{
  \begin{picture}(98,34) (143,-111)
    \SetWidth{1.0}
    \SetColor{Black}
    \GOval(192,-94)(16,16)(0){0.6}
    \Vertex(144,-94){2}
    \Line[arrow,arrowpos=0.5,arrowlength=5,arrowwidth=2,arrowinset=0.2](144,-94)(176,-94)
    \Line[arrow,arrowpos=0.5,arrowlength=5,arrowwidth=2,arrowinset=0.2](208,-94)(240,-94)
    \Vertex(240,-94){2}

  \end{picture}
}
\end{center}
and given precisely by the expression
\begin{eqnarray}
\mathcal{S}(k^{0},\vec{k})=\rmi \frac{k^{0}\gamma^{0}-Z(k)[\vec{k}.\vec{\gamma}-m(k)]}{(k^{0})^{2}-\omega_{0}^{2}(k)+\rmi\epsilon},\label{Chap2PropagatorCondensateTrue}
\end{eqnarray}
with $k=|\vec{k}|$ and
\begin{eqnarray}
Z(k)=\frac{1-\psi^{2}(k)}{1+\psi^{2}(k)}\frac{\omega_{0}(k)}{k}, \quad m(k)=\frac{2k\psi(k)}{1-\psi^{2}(k)}. 
\end{eqnarray}
The behaviour of the functions $m(k)$ and $Z(k)$ is illustrated in the figure (\ref{Chap2m_k}) and (\ref{Chap2Z_k}). As expected, the dynamical mass tends to zero at large momentum, while the function $Z(k)$ goes to unity. Whereas the value $m(0)$ is finite, we observe that $Z(k)$ exhibits an integrable logarithmic divergence as $k\to 0$.

\section{Correction to the magnetic mode propagator\label{SectionCorrectionMagneticMode}}
While the previous sections treated the fermion sector in the sole presence of Coulomb interactions, the present section aims at examining the influence of the dynamics of the fermions on the propagation of the magnetic mode. The non-perturbative solution in the fermionic sector will serve the zeroth-order contribution in the perturbative expansion in the interactions with the magnetic sector.

In order to understand the effect of the fermion condensate and of parity violation on the magnetic mode sector, it is instructive to review the one-loop correction to the photon propagator in the absence of a condensate, with a massless fermion. Indeed, UV divergences in perturbation theory, due to the large momentum regime, will affect the magnetic mode progator, irrespective of the presence or not of the condensate.

In relativistic covariant perturbation theory, the leading order correction is the amputated diagram

\begin{center}
\fcolorbox{white}{white}{
  \begin{picture}(200,67) (89,-109)
    \SetWidth{1.0}
    \SetColor{Black}
    \Text(85,-78)[]{$\rmi\Pi^{\mu\nu} = $}
    \Photon(123,-78)(154,-78){5.5}{3}
    \Arc[arrow,arrowpos=0.5,arrowlength=5,arrowwidth=2,arrowinset=0.2](176,-78)(22,225,585)
    \Vertex(198,-78){2}
    \Vertex(154,-78){2}
    \Photon(198,-78)(229,-78){5.5}{3}
  \end{picture}
}
\end{center}
which reads
\begin{eqnarray}
\rmi\Pi^{\mu\nu}(p)=-\int\frac{\rmd^{3}\ell}{(2\pi)^{3}}e^{2}{\rm Tr}[\gamma^{\mu}\frac{\slashed{p}+\slashed{\ell}}{(p+\ell)^{2}+\rmi\epsilon}\gamma^{\nu}\frac{\slashed{\ell}}{\ell^{2}+\rmi\epsilon}].
\end{eqnarray}
The integral is linearly divergent in power counting. While dimensional regularization provides a finite result without a divergent contribution \cite{JackiwTempleton}, we prefer here to use a cut-off regulator, because it is more instructive in this context, but at the expense of breaking gauge symmetry.
After a Wick rotation $\ell^{0}\to\rmi \ell^{0}_{E}$, and with the help of the Feynman parameter trick, we obtain the result
\begin{eqnarray}
\rmi\Pi^{\mu\nu}(p)=-\frac{\rmi e^{2}}{3\pi^{2}}\Lambda\eta^{\mu\nu}-\frac{\rmi e^{2}}{16}(\eta^{\mu\nu} p^{2}-p^{\mu}p^{\nu})\frac{1}{\sqrt{-p^{2}-\rmi \epsilon}},\label{Chap2CovariantPhotonPolarization}
\end{eqnarray}
where the linearly divergent contribution in the first term is a gauge symmetry breaking term, whereas the second term is the finite result also given by the dimensional regularization procedure\footnote{Here, $p=(p^{0},\vec{p})$ is the $3$-vector associated to the momentum of the incoming photon.}. The cut-off dependent term has to be subtracted exactly thanks to a covariant mass counter term in the Lagrangian, leaving no ambiguous finite term in order to preserve the Ward identity. 

As a lesson from the form of vacuum polarization contribution in the absence of the condensate, we expect also a linear divergence in the analogue diagram for the magnetic mode in the condensate.
Namely, we are interested in the two-point function of the magnetic mode
\begin{eqnarray}
T\langle \Omega|\Phi(x^{0},\vec{x})\Phi(y^{0},\vec{y})|\Omega\rangle
\end{eqnarray}
with $|\Omega\rangle$ the full interacting vacuum.
Working in $p$-space, we decide to perform a perturbative expansion, with the interaction Hamiltonians $\hat{H}^{I}_{\Phi\chi}$ and $\hat{H}^{I}_{C}$.
Hence the Feynman rule for the vertex between the magnetic mode and the current

\begin{center}
\fcolorbox{white}{white}{
  \begin{picture}(50,36) (79,-141)
    \SetWidth{1.0}
    \SetColor{Black}
    \Gluon(80,-124)(112,-124){1.5}{3}
    \Line[arrow,arrowpos=0.5,arrowlength=5,arrowwidth=2,arrowinset=0.2](112,-124)(128,-108)
    \Line[arrow,arrowpos=0.5,arrowlength=5,arrowwidth=2,arrowinset=0.2](112,-124)(128,-140)
    \Line[arrow,arrowpos=1,arrowlength=5,arrowwidth=2,arrowinset=0.2](80,-108)(96,-108)
    \Vertex(112,-124){2}
  \end{picture}
}
\end{center}
is given by: $e \epsilon^{ij}p^{j}\gamma^{j}$, when the magnetic mode momentum $(p^{0},\vec{p})$ is incoming.

With the help of this Feynman rule, it is possible to formulate the first loop correction to the free propagator.
In the presence of the condensate, the first contribution to vacuum polarization is

\begin{center}
\fcolorbox{white}{white}{
  \begin{picture}(154,67) (89,-109)
    \SetWidth{1.0}
    \SetColor{Black}
    \Text(90,-78)[]{$-\rmi\pi(p^{0},\vec{p})=$}
     \Line[arrow,arrowpos=1,arrowlength=5,arrowwidth=2,arrowinset=0.2](133,-68)(143,-68)
     \Line[arrow,arrowpos=1,arrowlength=5,arrowwidth=2,arrowinset=0.2](201,-68)(211,-68)
    \Gluon(130,-78)(154,-78){1.5}{4}
    \Arc[arrow,arrowpos=0.5,arrowlength=5,arrowwidth=2,arrowinset=0.2](176,-78)(22,225,585)
    \Vertex(198,-78){2}
    \Vertex(154,-78){2}
    \Gluon(198,-78)(222,-78){1.5}{4}
    \GOval(176,-58)(8,8)(0){0.6}
    \GOval(176,-98)(8,8)(0){0.6}
  \end{picture}
}
\end{center}
where the fermion propagator in the condensate is the one given by (\ref{Chap2PropagatorCondensateTrue}).
The polarization modified by the presence of the condensate can be written as the product
\begin{eqnarray}
\pi(p^{0},\vec{p})=|\vec{p}|^{2}\pi_{0}(p^{0},\vec{p}),
\end{eqnarray}
where the quantity of interest is the loop integral
\begin{eqnarray}
-\rmi \pi_{0}(p^{0},\vec{p})=-\int\frac{\rmd^{3}\ell}{(2\pi)^{3}}2e^{2}\times\Big\{\label{Chap2pi0}\\
\frac{\ell^{0}(p^{0}+\ell^{0})+Z(\ell)Z(|\vec{p}+\vec{\ell}|)[2\ell^{2}\sin^{2}\theta-\vec{\ell}.(\vec{\ell}+\vec{p})-m(\ell)m(|\vec{p}+\vec{\ell}|)]}{[(p^{0}+\ell^{0})^{2}-\omega^{2}(|\vec{p}+\vec{\ell}|)+\rmi \epsilon][(\ell^{0})^{2}-\omega^{2}(\ell)+\rmi \epsilon]}\Big\}\nonumber
\end{eqnarray}
with $\ell=|\vec{\ell}|$ and where $\theta$ is the relative angle between the loop momentum $\vec{\ell}$ and the incoming spatial momentum $\vec{p}$.
Denoting the free magnetic mode propagator by 
\begin{eqnarray}
D(p^{0},\vec{p})=\frac{1}{|\vec{p}|^{2}}\frac{\rmi}{(p^{0})^{2}-|\vec{p}|^{2}+\rmi\epsilon},
\end{eqnarray}
we can compute the full propagator as the sum of the one particle irreducible diagrams
\begin{eqnarray}
&D(p^{0},\vec{p})+(D(-\rmi\pi) D)(p^{0},\vec{p})+(D(-\rmi\pi) D(-\rmi\pi) D)(p^{0},\vec{p})+\dots\nonumber\\
&=\rmi\Big\{|\vec{p}|^{2}\big[(p^{0})^{2}-|\vec{p}|^{2}-\pi_{0}(p^{0},\vec{p})+\rmi\epsilon\big]\Big\}^{-1}.
\end{eqnarray}
Hence the investigation for a dynamical mass of the magnetic mode photon requires to solve the condition 
\begin{eqnarray}
(p^{0})^{2}-|\vec{p}|^{2}-\pi_{0}(p^{0},\vec{p})=0,\label{Chap2pole_equation}
\end{eqnarray}
in order to find the position of a pole of order one in the resummed propagator. If we can find a solution to (\ref{Chap2pole_equation}) in perturbation theory, we will be able to write a dispersion relation $p^{0}(|\vec{p}|)$, and will define a running mass squared as
\begin{eqnarray}
M^{2}(|\vec{p}|)=\Big(p^{0}(|\vec{p}|)\Big)^{2}-|\vec{p}|^{2}.
\end{eqnarray}
As we will show, the solution verifies, to leading order in perturbation theory,
\begin{eqnarray}
(p^{0})^{2}=|\vec{p}|^{2}+e^{4}\pi'(p^{0},\vec{p})\approx |\vec{p}|^{2}+e^{4}\pi'(|\vec{p}|,\vec{p}),
\end{eqnarray}
where we used $\pi_{0}(p^{0},\vec{p})=e^{4}\pi'(p^{0},\vec{p})$, so that the running mass squared is approximately given by
\begin{eqnarray}
M^{2}(|\vec{p}|)=\Big(p^{0}(|\vec{p}|)\Big)^{2}-|\vec{p}|^{2}\approx e^{4}\pi'(|\vec{p}|,\vec{p}).
\end{eqnarray}
The value that we will be interested in, is $M^{2}(0)\approx e^{4}\pi'(0,\vec{0})$. Hence, due to the technical difficulties, we shall only calculate the value of $\pi_{0}(0,\vec{0})$.

Because the computation of $\pi_{0}(p^{0},\vec{p})$ involves a function known only numerically, we shall evaluate its first term in a power expansion in $|\vec{p}|$,
\begin{eqnarray}
\pi(p^{0},\vec{p})=|\vec{p}|^{2}\Big[\pi_{0}(p^{0},\vec{0})+O(|\vec{p}|)\Big].
\end{eqnarray}
The expression of $\pi_{0}(p^{0},\vec{0})$ in (\ref{Chap2pi0}) involves an integral over the temporal and spatial loop momentum of a non explicitly covariant function, so that Wick rotation does not seem to be appropriate. Nevertheless the Feynman parameter technique can be used and, afterwards, the expression can be simplified thanks to the shift $\ell^{0}\to \ell^{0}-xp^{0}$. The $\ell^{0}$-integral is convergent and can be calculated by evaluating the residue of a double pole, leaving an integral over the spatial momentum $\vec{\ell}$. Performing the angular integral, the result is an integral over $\ell=|\vec{\ell}|$,
\begin{eqnarray}
-\rmi\pi_{0}^{\Lambda}(p^{0},\vec{0})=\frac{-\rmi e^{2}}{2}\int_{0}^{1}\rmd x\int_{0}^{\Lambda}\frac{\ell\rmd \ell}{2\pi} \frac{-Z^{2}(\ell)\ell^{2}-2Z^{2}(\ell)m^{2}(\ell)}{[\omega_{0}^{2}(\ell)-x(1-x)(p^{0})^{2}]^{3/2}},
\end{eqnarray}
whose linear divergence was regularised with a cut-off $|\vec{\ell}|<\Lambda$. The divergent behaviour lies of course in the ultraviolet regime and is exactly the same as in the absence of a condensate.  Neglecting the condensate, that is to say putting $\Psi=0$, we find the exact result
\begin{eqnarray}
-\rmi|\vec{p}|^{2}\pi^{\Lambda}_{0}(p^{0},\vec{0})&\stackrel{\Psi= 0}{=}&|\vec{p}|^{2}\frac{-\rmi e^{2}}{2}\int_{0}^{1}\rmd x\int_{0}^{\Lambda}\frac{\ell\rmd \ell}{2\pi} \frac{-\ell^{2}}{[\ell^{2}-x(1-x)(p^{0})^{2}]^{3/2}},\nonumber\\
&=&|\vec{p}|^{2} \frac{-\rmi e^{2} \sqrt{-(p^{0})^{2}}}{16}+|\vec{p}|^{2}\frac{\rmi e^{2}}{2}\frac{\Lambda}{2\pi}.\label{Chap2NonCovariantPhotonPolarization}
\end{eqnarray}
 Because we expect that, in the large momentum limit, the theory with the condensate yields the same result as ordinary perturbative \cal{QED}$_{2+1}$, the requirement of finiteness of this diagram gives us an unambiguous way to subtract the linear divergence of the same diagram in presence of the condensate. Hence, using (\ref{Chap2NonCovariantPhotonPolarization}),  the renormalization of $\pi_{0}(p^{0},\vec{0})$ gives a finite result
\begin{eqnarray}
-\rmi\pi_{0}^{{\rm reg}}(p^{0},\vec{0})=\lim_{\Lambda\to + \infty}\Big\{ -\rmi \pi_{0}^{\Lambda}(p^{0},\vec{0})-\frac{\rmi e^{2}}{2}\frac{\Lambda}{2\pi} \Big\},
\end{eqnarray}
obtained thanks to the addition of a counter term proportional to $\Phi\Delta\Phi$ in the Lagrangian. Setting $p^{0}=0$ in order to evaluate the mass of the magnetic mode, a numerical integration yields the result
\begin{eqnarray}
\pi_{0}^{{\rm reg}}(0,\vec{0})&=&-\frac{e^{2}}{4\pi}\int_{0}^{+\infty}\rmd \ell\Big\{\frac{\ell-\omega(\ell)}{\omega(\ell)}+\frac{\ell Z^{2}(\ell)m^{2}(\ell)}{\omega^{3}(\ell)}\Big\}\nonumber\\
&\approx & 0.14\Big(\frac{e^{2}}{4\pi}\Big)^{2}.
\end{eqnarray}
The subtraction of the linear divergence from this one loop diagram leaves a finite contribution proportional to $e^{4}$. Other finite contributions proportional to $e^{4}$ will come from diagrams containing more loops. However, it is not excluded that two loop diagrams give rise to a divergent dynamical mass to the magnetic mode. Among them, the potentially problematic diagram denoted by $-\rmi \pi^{(1)}(p^{0},\vec{p})$, 

\begin{center}
\fcolorbox{white}{white}{
  \begin{picture}(310,67) (70,-109)
    \SetWidth{1.0}
    \SetColor{Black}
    \Text(100,-78)[]{$-\rmi\pi^{1}(p^{0},\vec{p}) = $}
    \Line[arrow,arrowpos=1,arrowlength=5,arrowwidth=2,arrowinset=0.2](137,-68)(147,-68)
    \Gluon(134,-78)(154,-78){1.5}{4}
    \Arc[arrow,arrowpos=0.5,arrowlength=5,arrowwidth=2,arrowinset=0.2](176,-78)(22,225,585)
    \Vertex(198,-78){2}
    \Vertex(154,-78){2}
    \GOval(176,-58)(8,8)(0){0.6}
    \GOval(176,-98)(8,8)(0){0.6}
    \Photon(198,-78)(259,-78){6.5}{5}
    \Arc[arrow,arrowpos=0.5,arrowlength=5,arrowwidth=2,arrowinset=0.2](281,-78)(22,225,585)
    \Vertex(259,-78){2}
    \Vertex(303,-78){2}
    \Gluon(303,-78)(323,-78){1.5}{4}
    \Line[arrow,arrowpos=1,arrowlength=5,arrowwidth=2,arrowinset=0.2](306,-68)(316,-68)
    \GOval(281,-58)(8,8)(0){0.6}
    \GOval(281,-98)(8,8)(0){0.6}
  \end{picture}
}
\end{center}
with an intermediate Coulomb propagator, could provide an additional contribution to the mass of the magnetic mode. Because of the intermediate Coulomb propagator $\rmi/ |\vec{p}|^{2}$, we could expect that $-\rmi \pi^{(1)}(0,\vec{0})={\rm Cst}\neq 0$, so that it gives rise to a pole in the dynamical mass as $|\vec{p}|\to 0$.
However, this is not the case. The diagram is of the form
\begin{eqnarray}
-\rmi \pi^{(1)}(p^{0},\vec{p})=(-\rmi\kappa(p^{0},\vec{p}))\frac{\rmi}{|\vec{p}|^{2}}(-\rmi\kappa(p^{0},\vec{p})),
\end{eqnarray}
where the first order in the expansion in $|\vec{p}|$ and $p^{0}$ can be found thanks to 
\begin{eqnarray}
\kappa(p^{0},\vec{p})\approx|\vec{p}|^{2}\frac{e^{2}}{4\pi}\kappa_{0},
\end{eqnarray}
with the numerical coefficient given by the quadrature
\begin{eqnarray}
\kappa_{0}=\int_{0}^{+\infty}\ell \rmd \ell \ Z^{2}(\ell)\frac{m(\ell)-\ell m'(\ell)/2}{\omega_{0}^{3}(\ell)}\approx 0.58.
\end{eqnarray}
Defining $\pi^{(1)}(p^{0},\vec{p})=|\vec{p}|^{2}\pi^{(1)}_{0}(p^{0},\vec{p})$, we find the contribution to the mass of this diagram to be
\begin{eqnarray}
\pi^{(1)}_{0}(0,\vec{0})\approx 0.34 \Big(\frac{e^{2}}{4\pi}\Big)^{2}.
\end{eqnarray}
We may find the approximate value of the mass of the magnetic mode by summing the contributions coming from the two diagrams considered, i.e. $M^{2}(0)\approx 0.48 (e^{2}/4\pi)^{2}$.
\section{Conclusion \label{Conclusion2+1}}
Thanks to the factorization of local gauge transformations and of gauge degrees of freedom, as well as the dressing of the fermion field, the dynamics of  massless \cal{QED}$_{2+1}$ with one flavour of electrons could be reduced to the interaction of a dressed fermion field with a physical magnetic scalar mode. The decomposition of the gauge field and the factorization of the local gauge symmetry rendered manifest the relevance of the gauge invariant magnetic scalar, understood as the only propagating gauge invariant electromagnetic degree of freedom.

In the fermionic sector, a ground state of the BCS type was shown to be energetically more favourable than an ``empty'' Fock state. Furthermore, the wave function of the pair condensate was found by solving an integral equation, including non-perturbatively the effects of Coulomb interactions. As a result, the pseudo-particle excitations above the condensate, namely the constituent fermions, exhibit a peculiar dispersion relation, with a divergent behaviour at low momentum, being a signature for the confinement of charged states. This interpretation was confirmed by the study of the energy of a bound state of two of these constituent fermions.

Due to pair condensation, parity symmetry is spontaneously broken. Hence, the propagation of the magnetic mode excitations is affected by the interactions with the pair condensate. Starting from the non-perturbative result for the ground state, we decided to expand in perturbation the effects of the residual Coulomb interactions and the interactions between the magnetic mode and the fermion current.
Although the complete loop calculation seems to be too involved, the corrections to the magnetic mode propagator from the first relevant diagrams indicate the dynamical generation of a mass for the magnetic mode.

Among the drawbacks of the variational approach used here, the difficulty to evaluate the accuracy of the implied approximation is a disadvantage. In contradistinction to a perturbative treatment,  no power expansion in a small parameter is performed to obtain the ground state. It is the form of the pair condensate state which dictates the form of the integral equation to be solved. Hence, in order to improve the reliability of the approximation, the flexibility of the ansatz wave function could be increased.
As a perspective, it would be instructive to study the possibility of a condensation of magnetic modes, in interaction with condensed fermion pairs. This idea has been explored in a recent work in the case of \cal{QCD}$_{3+1}$, in a ``quenched'' approximation of \cal{QCD}$_{3+1}$ \cite{PakReinhardt}.

Due to the factorization of the local gauge symmetry, the formulation used in this work has lost manifest Lorentz covariance, although it remains covariant under spatial translations and rotations. It is challenging to understand how the equations are changed under a Lorentz boost. We leave this analysis for a further work. Nevertheless, one conclusion seems to have been established definitely by the present work. The well-known exact solution to the Schwinger model, namely massless \cal{QED}$_{1+1}$, shows that as soon as the gauge coupling constant is turned on however small its value, massless quantum electrodynamics in two spacetime dimensions is not a theory of interacting (and gauge non invariant) electrons and photons, but rather is a theory of a (gauge invariant) free massive pseudoscalar particle, namely essentially the electric field. Likewise massless quantum electrodynamics in three spacetime dimensions with a non vanishing gauge coupling constant however small its value, is not a theory of interacting (and gauge non invariant) electrons and photons, but rather is a theory of a (gauge invariant) massive magnetic mode scalar interacting with (gauge invariant) neutral paired electron-positron states. Furthermore, parity is spontaneously broken dynamically, while charged states cannot be separed at large distances and remain confined in the neutral paired electron-positron states.
\ack

It is a pleasure to acknowledge Gauthier Durieux, Philippe Mertens and Mathieu Buchkremer for stimulating discussions. The work of MF was supported by the National Fund for Scientific Research
(F.R.S.-FNRS, Belgium) through a ``Aspirant'' Research fellowship. This work is supported by the Belgian Federal Office
for Scientific, Technical and Cultural Affairs through the Interuniversity Attraction Pole P6/11.

\appendix

\section{The Hadamard finite part and the photon mass term\label{HadamardAnnexe}}
\begin{figure}[!ht]
\centering
\includegraphics{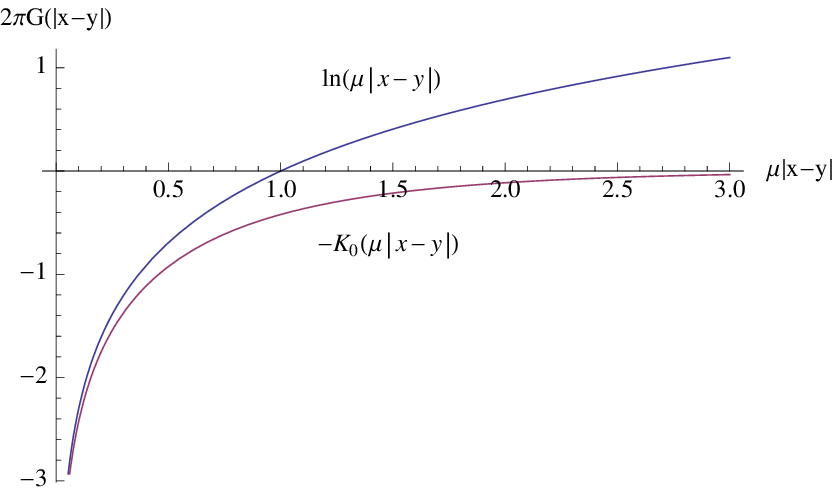}
\caption{The figure compares the behaviour of the $x$-space Green function in presence and absence of a mass term for the photon. The large distance behaviours are very different. \label{FigureGrennFunction}} 
\end{figure}

The Fourier transform of the $x$-space Green function is not a function but a distribution. It may be more convincing to obtain the Hadamard finite part in terms of a limiting case of a more intuitive situation.
The naive $\frac{-1}{|\vec{p}|^{2}}$ infrared divergent  $p$-space Green function can be regularised using a mass regulator.
If one adds a mass term in the Green function in  $p$-space, one finds the following $x$-space Green function
\begin{eqnarray}
G_{\mu}(x,y)=\int\frac{\rmd^{2}p^{i}}{(2\pi)^{2}}\frac{-1}{|\vec{p}|^{2}+\mu^{2}}e^{\rmi\vec{p}.(\vec{x}-\vec{y})}=-\frac{1}{2\pi}K_{0}(\mu|\vec{x}-\vec{y}|),\label{BesselK}
\end{eqnarray}
where $K_{0}(\mu|\vec{x}-\vec{y}|)$ is a modified Bessel function of the second kind.
The IR behaviour of $G_{\mu}(x,y)$ completely changes however small the value for $\mu$ is, as illustrated in Fig. \ref{FigureGrennFunction}. Even for a very small $\mu$, the ``potential'' $G_{\mu}(x,y)$ is no longer confining! 

A brutal substitution $\mu= 0$ in the last Fourier transform gives us the naive Fourier transform of the Green function.
However we know that the limit $\mu\to 0$ should be taken with care. Setting $\mu= 0$ barely makes sense. The reason for this is that when $\mu$ goes to zero, the integration in (\ref{BesselK}) still involves values of $\vec{p}$ with $|\vec{p}|<\mu$. In order to identify the divergence resulting from the limit $\mu\to 0$, one may clearly separate the safe regions of integration from the potentially divergent regions. To do so, one introduces $\epsilon>\mu$, which will be kept constant in the limit $\mu\to 0$. Hence we can rewrite
\begin{eqnarray}
G_{\mu}(x,y)=I_{1}^{\epsilon}+I_{2}^{\epsilon},
\end{eqnarray}
with
\begin{eqnarray}
&I_{1}^{\epsilon}=\int_{0}^{\epsilon}\frac{p\rmd p}{2\pi}\frac{-1}{p^{2}+\mu^{2}}J_{0}(p|\vec{x}-\vec{y}|),\label{FirstTermMu}\\
&I_{2}^{\epsilon}=\int_{\epsilon}^{\infty}\frac{p\rmd p}{2\pi}\frac{-1}{p^{2}+\mu^{2}}J_{0}(p|\vec{x}-\vec{y}|).
\end{eqnarray}
It is now straightforward to take the limit of the second term
\begin{eqnarray}
\lim_{\mu\to 0}I_{2}^{\epsilon}=\int_{\epsilon}^{\infty}\frac{\rmd p}{2\pi}\frac{-1}{p}J_{0}(p|\vec{x}-\vec{y}|),
\end{eqnarray}
where $ J_{0}$ is a Bessel function of the first kind.
One may also consider the first term and extract its divergent contribution when $\mu\to 0$. Integrating it by parts one finds
\begin{eqnarray}
&I_{1}^{\epsilon}=&\frac{1}{2\pi}(-\frac{1}{2}\ln\frac{\epsilon^{2}+\mu^{2}}{\epsilon^{2}}J_{0}(\epsilon|\vec{x}-\vec{y}|)+\ln\frac{\mu}{\epsilon})+\\
&&+\int_{0}^{\epsilon}\frac{\rmd p}{2\pi}\frac{1}{2}\ln(\frac{p^{2}+\mu^{2}}{\epsilon^{2}})|\vec{x}-\vec{y}|J_{1}(p|\vec{x}-\vec{y}|),
\end{eqnarray}
where, as before $ J_{1}$ denotes a Bessel function of the first kind.
The second term in the last equation is perfectly convergent when $\mu\to 0$. We have succeeded in pinpointing the divergent contribution occuring when the mass goes to zero.
It is now completely obvious that the behaviour of $I_{1}^{\epsilon}$ in the limit is 
\begin{eqnarray}
&\lim_{\mu\to 0}I_{1}^{\epsilon}=\lim_{\mu\to 0}\frac{1}{2\pi}\ln\frac{\mu}{\epsilon}+\int_{0}^{\epsilon}\frac{\rmd p}{2\pi}\frac{1}{2}\ln(\frac{p^{2}}{\epsilon^{2}})|\vec{x}-\vec{y}|J_{1}(p|\vec{x}-\vec{y}|).
\end{eqnarray}
The only source of divergence is the term $\frac{1}{2\pi}\ln\frac{\mu}{\epsilon}$ that needs to be subtracted from $I_{1}^{\epsilon}$ to make sense of the limit. One notices also that the quantity that has to be added to $I_{1}^{\epsilon}$ to ensure the subtraction is
\begin{eqnarray}
-\frac{1}{2\pi}\ln\frac{\mu}{\epsilon}=\int_{\mu}^{\epsilon}\frac{\rmd p}{2\pi}\frac{1}{p}=\int_{0}^{\epsilon}\frac{\rmd p}{2\pi}\frac{1}{p}\theta(p-\mu).
\end{eqnarray}
Adding this term to (\ref{FirstTermMu}), and taking the limit, one finds
\begin{eqnarray}
&\lim_{\mu\to 0}I_{1}^{\epsilon}-\frac{1}{2\pi}\ln\frac{\mu}{\epsilon}&=\lim_{\mu\to 0}\int_{0}^{\epsilon}\Big\{\frac{\rmd p}{2\pi}\frac{-p}{p^{2}+\mu^{2}}J_{0}(p|\vec{x}-\vec{y}|)+\frac{1}{p}\theta(p-\mu)\Big\}\nonumber\\
&&=\int_{0}^{\epsilon}\frac{\rmd p}{2\pi}[\frac{-1}{p}J_{0}(p|\vec{x}-\vec{y}|)+\frac{1}{p}]
\end{eqnarray}
Restoring now the angular integral by replacing the Bessel function by its integral representation, the final result of this procedure is 
\begin{eqnarray}
\lefteqn{\lim_{\mu\to 0}G_{\mu}(x,y)-\frac{1}{2\pi}\ln\frac{\mu}{\epsilon}}\\
&=\int_{|\vec{p}|<\epsilon}\frac{\rmd^{2}p^{i}}{(2\pi)^{2}}\frac{-1}{|\vec{p}|^{2}}(e^{\rmi\vec{p}.(\vec{x}-\vec{y})}-1)+\int_{|\vec{p}|>\epsilon}\frac{\rmd^{2}p^{i}}{(2\pi)^{2}}\frac{-1}{|\vec{p}|^{2}}e^{\rmi\vec{p}.(\vec{x}-\vec{y})}.
\end{eqnarray}
Hence in conclusion, the Hadamard finite part can indeed be interpreted as the limit of the Green function regularised with a mass term for the photon.
The presence of the scale $\epsilon$ is unavoidable because it is essential to help us to make sense of the limit $\mu\to 0$ which is a limit of a dimensionful quantity.
The scale $\epsilon$ is somehow a remnant of the mass term.
\section{Matrix elements and contractions \label{MatrixElements}}
Some useful matrix elements are
\begin{eqnarray}
&\langle \Psi|\chi_{\alpha}^{\dagger}(0,\vec{x})\chi_{\beta}(0,\vec{y})|\Psi\rangle
= \int\frac{\rmd^{2} p^{i}}{2p^{0}}\Big[p^{0}-(1-2|\beta(p)|^{2})\gamma^{0}\vec{\gamma}.\vec{p}\label{MatrixElement1}\\&
-p^{0}\gamma^{0}\alpha(p)[\beta(p)+\beta^{*}(p)]+\vec{p}.\vec{\gamma}\alpha(p)[\beta(p)-\beta^{*}(p)]\Big]_{\beta\alpha}\frac{e^{-\rmi \vec{p}(\vec{x}-\vec{y})}}{(2\pi)^{2}},\nonumber\\
& \langle \Psi|\chi_{\alpha}(0,\vec{x})\chi_{\beta}^{\dagger}(0,\vec{y})|\Psi\rangle=\int\frac{\rmd^{2} p^{i}}{2p^{0}}\Big[p^{0}+(1-2|\beta(p)|^{2})\gamma^{0}\vec{\gamma}.\vec{p}\nonumber\\&
+p^{0}\gamma^{0}\alpha(p)[\beta(p)+\beta^{*}(p)]-\vec{p}.\vec{\gamma}\alpha(p)[\beta(p)-\beta^{*}(p)]\Big]_{\alpha\beta}\frac{e^{\rmi \vec{p}(\vec{x}-\vec{y})}}{(2\pi)^{2}}.\label{MatrixElement2}
\end{eqnarray}
The contractions needed to compute the matrix elements of the normal ordered operators are
\begin{eqnarray}
&\widehat{\chi_{\alpha}^{\dagger}(0,\vec{x})\chi_{\beta}(0,\vec{y})}= \int\frac{\rmd^{2} p^{i}}{2p^{0}}\Big[2|\beta(p)|^{2}\gamma^{0}\vec{\gamma}.\vec{p}\\&
-p^{0}\gamma^{0}\alpha(p)[\beta(p)+\beta^{*}(p)]+\vec{p}.\vec{\gamma}\alpha(p)[\beta(p)-\beta^{*}(p)]\Big]_{\beta\alpha}\frac{e^{-\rmi \vec{p}(\vec{x}-\vec{y})}}{(2\pi)^{2}},\nonumber\\
&\widehat{\chi_{\alpha}(0,\vec{x})\chi^{\dagger}_{\beta}(0,\vec{y})}=\int\frac{\rmd^{2} p^{i}}{2p^{0}}\Big[-2|\beta(p)|^{2}\gamma^{0}\vec{\gamma}.\vec{p}\\&
+p^{0}\gamma^{0}\alpha(p)[\beta(p)+\beta^{*}(p)]-\vec{p}.\vec{\gamma}\alpha(p)[\beta(p)-\beta^{*}(p)]\Big]_{\alpha\beta}\frac{e^{\rmi \vec{p}(\vec{x}-\vec{y})}}{(2\pi)^{2}}.\nonumber
\end{eqnarray}

\section{Useful Integrals \label{UsefulIntegrals}}
The following integrals have to be computed with great care:
\begin{eqnarray}
&\int  \frac{\rmd \theta}{p^{2}+q^{2}-2 pq \cos \theta}=\frac{2}{|p^{2}-q^{2}|} \rm{Atan}\big\{\frac{p+q}{|p-q|}\tan \theta/2\big\},\\
&\int \frac{\cos \theta\rmd \theta}{p^{2}+q^{2}-2 pq \cos \theta}=\frac{1}{2pq}\Big\{-\theta+2\frac{p^{2}+q^{2}}{|p^{2}-q^{2}|} \rm{Atan}[\frac{p+q}{|p-q|}\tan \frac{\theta}{2}]\Big\},\\
&\int_{0}^{2\pi}  \frac{\rmd \theta}{p^{2}+q^{2}-2 pq \cos \theta}=\frac{2\pi}{|p^{2}-q^{2}|},\\
&\int_{0}^{2\pi} \frac{\cos \theta\rmd \theta}{p^{2}+q^{2}-2 pq \cos \theta}=\frac{2\pi}{2pq}\Big\{-1+\frac{p^{2}+q^{2}}{|p^{2}-q^{2}|}\Big\},
\end{eqnarray}
where the evaluation of the definite integrals takes into account the presence of a discontinuity in the corresponding primitives.

\section{The self-energy contribution to the dispersion relation \label{Self-EnergySigma}}

At equation (\ref{Chap2RegularizedSmallp}) we found an interesting result and provide here some details for its derivation. We had to evaluate the finite part of the problematic integral
\begin{eqnarray}
\lefteqn{\sigma(p)=\frac{e^{2}}{2}\mathcal{P}\int \frac{\rmd ^{2} q^{i}}{(2\pi)^{2}}\frac{\vec{p}.\vec{q}}{|\vec{p}||\vec{q}|}\frac{1}{(\vec{p}-\vec{q})^{2}}}\\
&=\frac{e^{2}}{2(2\pi)^{2}}\int_{0}^{+\infty}\rmd q \frac{1}{q}\Big[ \int_{0}^{2\pi}\rmd \theta \frac{p+q\cos\theta}{\sqrt{p^{2}+q^{2}+2pq\cos\theta}}-2\pi H(\mu-q)\Big]
\end{eqnarray}
where $H(x)$ is the Heaviside step function. Using 
\begin{eqnarray}
\frac{\partial}{ \partial q}(\frac{p+q\cos\theta}{\sqrt{p^{2}+q^{2}+2pq\cos\theta}})=\frac{-p q\sin^{2} \theta }{(p^{2}+q^{2}+2pq\cos\theta)^{3/2}}
\end{eqnarray}
and an integration by parts (with vanishing boundary terms), we find
\begin{eqnarray}
\sigma(p)=\frac{e^{2}}{4\pi}\ln\frac{c}{\mu}+\frac{e^{2}}{8\pi^{2}}\int_{0}^{+\infty}\rmd q \ \ln\frac{q}{c}\int_{0}^{2\pi}\rmd\theta \frac{pq \sin^{2}\theta}{(p^{2}+q^{2}+2pq\cos\theta)^{3/2}}
\end{eqnarray}
where $c$ is an integration constant. One can first perform a change of variables $q=p s$ and then calculate the $s$-integral. The final result is a function of $\theta$, which can be integrated from $0$ to $2\pi$.
The integration constant simplifies, and the result is 
\begin{eqnarray}
\sigma(p)=\frac{e^{2}}{4\pi}[\ln(\frac{2p}{\mu})+\ln2-1].
\end{eqnarray}

\section{Feynman Rules\label{FeynmanRulesAnnexe}}
The Feynman rules associated to the Schwinger-Dyson equations of Section \ref{Schwinger_Dyson} are:

\begin{center}
\fcolorbox{white}{white}{
  \begin{picture}(280,271) (41,-55)
    \SetWidth{1.0}
    \SetColor{Black}
    \GOval(114,195)(19,19)(0){0.882}
    \Vertex(57,195){2}
    \Line[arrow,arrowpos=0.5,arrowlength=5,arrowwidth=2,arrowinset=0.2](57,195)(95,195)
    \Line[arrow,arrowpos=0.5,arrowlength=5,arrowwidth=2,arrowinset=0.2](133,195)(171,195)
    \Vertex(171,195){2}
    
    \Vertex(57,138){2}
    \Line[arrow,arrowpos=0.5,arrowlength=5,arrowwidth=2,arrowinset=0.2](57,138)(171,138)
    \Vertex(171,138){2}
    \Line[arrow,arrowpos=0.5,arrowlength=5,arrowwidth=2,arrowinset=0.2](57,81)(95,81)
    \COval(114,81)(19,19)(0){Black}{White}
    \Line[arrow,arrowpos=0.5,arrowlength=5,arrowwidth=2,arrowinset=0.2](133,81)(171,81)
    \Vertex(57,24){2}\Photon(57,24)(171,24){7.5}{6}
    \Vertex(171,24){2}
    \Line[arrow,arrowpos=0.5,arrowlength=5,arrowwidth=2,arrowinset=0.2](57,-52)(114,-52)
    \Line[arrow,arrowpos=0.5,arrowlength=5,arrowwidth=2,arrowinset=0.2](114,-52)(171,-52)
    \Photon(114,-14)(114,-52){7.5}{2}
    \Vertex(114,-52){2}
    \Text(38,195)[lb]{\Black{$\alpha$}}
    \Text(190,195)[lb]{\Black{$\beta$}}
    \Text(38,138)[lb]{\Black{$\alpha$}}
    \Text(190,138)[lb]{\Black{$\beta$}}
    \Text(38,81)[lb]{\Black{$\alpha$}}
    \Text(190,81)[lb]{\Black{$\beta$}}
    \Text(38,-52)[lb]{\Black{$\alpha$}}
    \Text(190,-52)[lb]{\Black{$\beta$}}
    \Text(266,195)[lb]{\Black{$=S^{(3)}_{\alpha\beta}(p)$}}
    \Text(266,138)[lb]{\Black{$=S_{0}^{(3)}(p)_{\alpha\beta}$}}
    \Text(266,81)[lb]{\Black{$=-\rmi\Sigma_{\alpha\beta}(\vec{p})$}}
    \Text(266,24)[lb]{\Black{$=\rmi/|\vec{q}|^{2}$}}
    \Text(266,-52)[lb]{\Black{$=\rmi e (\gamma^{0})_{\alpha\beta}$}}

  \end{picture}
}
\end{center}
where 
\begin{eqnarray}
&S^{(3)}(p^{0},\vec{p})&=\frac{\rmi}{\slashed{p}-\Sigma(p^{0},\vec{p})+\rmi \epsilon},\\
&S^{(3)}_{0}(p^{0},\vec{p})&=\frac{\rmi}{\slashed{p}+\rmi \epsilon},\\
&\Sigma(p^{0},\vec{p})&=|\vec{p}|A(|\vec{p}|)+\vec{p}.\vec{\gamma} B(|\vec{p}|).
\end{eqnarray}

\section*{References}
\bibliography{biblio}
\bibliographystyle{hunsrt}
\end{document}